\documentclass[aps,prd,reprint,superscriptaddress,nofootinbib]{revtex4-2}
\usepackage{amsmath,amssymb,amsthm}
\usepackage{graphicx}
\usepackage{xcolor}
\usepackage{hyperref}
\hypersetup{colorlinks,linkcolor=blue!60!black,citecolor=blue!60!black,urlcolor=blue!60!black}

\newcommand{\rec}{\mathcal{R}}          
\newcommand{\Dis}{\mathcal{D}}          
\newcommand{\Edis}{E_{\mathrm{dis}}}
\newcommand{\EB}{E_{B}}
\newcommand{\Eesc}{E_{\mathrm{esc}}}
\newcommand{\Enu}{E_{\nu}}

\newcommand{\floor}[1]{\sigma_{\star}^{2}\!\left(#1\right)}
\newcommand{\fem}{f_{\mathrm{em}}}
\newcommand{\CRB}{\mathrm{CRB}}
\newcommand{\MMSE}{\mathrm{MMSE}}
\newcommand{\Ifish}{\mathcal{I}}

\newtheorem{prop}{Proposition}
\newtheorem{lem}{Lemma}

\begin{document}

\title{Calorimetry as inference: the information content of particle showers\\
and the limits of energy measurement}

\author{Wonyong Chung}
\affiliation{Department of Physics, Princeton University, Princeton, New Jersey 08544, USA}

\date{\today}

\begin{abstract}
A calorimeter does not measure the energy of a particle, it infers it from the record
the absorption process leaves behind. Energy conservation writes every event as an
accounting ledger, $E = \Edis + \EB + \Eesc + \Enu$, of which only the dissipated
component contributes to signal, so resolution is limited by how well the record of the
dissipation constrains the missing entries (binding, leakage, neutrinos). We formalize
this: records form a lattice on which information is monotone under refinement;
independent thinnings of a Poisson record (sampling fraction, photostatistics) scale
Fisher information exactly by the retention; summed readout of clustered records costs
the $1/(1+\mathrm{cv}^2)$ of classic sampling fluctuations; the quadrature budget
$a/\sqrt{E}\oplus b$ is the achievable floor of every summed readout and the
Gaussian-equality case of a two-sided information bound; and loss channels independent
of the record set hard floors that survive, in weakened form and under stated
hypotheses, the physical mixed case in which many events carry no loss. Within this
framework the ``ultimate limits'' of hadronic calorimetry are identified as the first
rung of a ladder, i.e. as the residual spread of the invisible energy around a
\emph{single} correlated observable. We measure the ladder above that rung on Geant4
truth-level records of hadronic showers in copper -- from the dissipated-energy sum
through dual-readout, to time-, position- and multiplicity-resolved records -- with
trained estimators validated on exactly solvable cascades. The achieved full-record
value, model-conditional, is approximately $5.5\%/\sqrt{E} \oplus 0.3\%$ over
20--100 GeV, a factor $1.7$--$1.9$ below the canonical one-functional construction
evaluated on the same simulated events with identical energy accounting ($1.29\%$ vs.
$2.14\%$ at 20 GeV; the originally published value is $2.7\%$). Retrained under
high-precision neutron transport, the full-record value moves from $1.29\%$ to
$1.30\%$ at 20 GeV, the dissipated-energy sum by $7\%$, and the resolution gained by
adding time structure becomes $0.17\%$ in place of $0.28\%$ (in absolute $\sigma/E$).
Dual-readout supplies the largest single share of the information, followed by the
shower shape, whose usable content is almost entirely longitudinal and saturates at a
sampling scale of about one interaction length. Finer segmentation, down to half an
interaction length, added nothing measurable for single-hadron energy. Time structure
improves the resolution by nearly the same amount whether or not shower shape
information is present, and stopped-muon neutrino losses set an estimated floor of $0.99\%$ at
20 GeV. We show that every readout dimension acquires a measured information exchange
rate -- the price of a decade of gate length, a $\lambda_I$ of cell size, a decade of
light yield -- separating what new hardware can buy from what better algorithms can
recover.
\end{abstract}

\maketitle

\section{Introduction}\label{sec:intro}

A calorimeter is conventionally described as a device that measures energy by total
absorption. Strictly speaking, it does no such thing. Energy conservation writes every
absorption event as an accounting ledger,
\begin{equation}
  E \;=\; \Edis \;+\; \EB \;+\; \Eesc \;+\; \Enu,
  \label{eq:ledger}
\end{equation}
in which only the first entry, energy dissipated into transducible channels (e.g.
ionization, scintillation- and Cherenkov-capable excitation, heat), ever contributes to
signal. Nuclear binding-energy loss $\EB$ vanishes into the mass excess of rearranged
nuclei, $\Eesc$ leaves the instrumented volume or arrives after the readout gate, and
$\Enu$ is carried away by neutrinos. A calorimeter therefore \emph{infers} the incident
energy from the record its absorber material keeps of the dissipation, and its
resolution is limited, not metaphorically but in the precise sense of estimation
theory, by how well that record constrains the missing elements of the ledger. The
classic fluctuation analyses~\cite{Amaldi1981,FabjanGianotti2003} enumerated the
missing entries, and here we quantify them. Every technique in the field's repertoire
is both a choice of the record and a statistic of it. The classical sum uses one
number: compensating calorimeters~\cite{WigmansRMP2018} up-weight neutron-rich
signals in hardware, dual-readout~\cite{WigmansRMP2018,Eno2025DR} measures two
functionals and solves for the electromagnetic fraction, software
compensation~\cite{CALICE2012SC} re-weights a segmented record; modern neural networks
consume the whole record, wholesale.

The community's canonical statement of the ``ultimate'' hadronic resolution is the
analysis of Lee, Livan and Wigmans~\cite{LeeLivanWigmans2018}: the spread of the
invisible energy that remains after conditioning on a \emph{single} correlated
observable, e.g. the electromagnetic fraction for dual-readout or the kinetic neutron
energy for compensation, evaluated with Geant4 truth-level accounting in massive
absorber blocks. Indeed, that very analysis shows that this is not a floor. Adding the neutron
time-structure fraction $f_n$ as a second conditioning variable improves upon it (the
DREAM data of Ref.~\cite{Akchurin2009Neutron}, reproduced as Fig.~11 of
Ref.~\cite{LeeLivanWigmans2018}), and reconstruction studies since have kept climbing.
Software compensation~\cite{CALICE2012SC}, interaction-vertex
imaging~\cite{VertexImaging2024}, neural-network regressions against calorimeter
granularity and timing~\cite{Neubuser2022,Akchurin2021NN}, and full-record
machine-learning regressions at low energy~\cite{Fudan2026ML} each extract more, and
networks trained as approximations to optimal reconstruction have approached the
record's limit in some cases. What has not yet been discussed in the literature is the
object these studies all implicitly approach: the information-theoretic limit of
the record itself, its structural properties on the lattice of records, or its value at
collider energies. Fisher information has long been used as a design tool for gamma-ray
detectors and their energy estimators~\cite{Bora2015,Salcin2014}, while
mutual-information objectives have been used to optimize calorimeter
layers~\cite{Wozniak2025}; and the hadronic energy ledger has its own
literature~\cite{Groom2007,Gabriel1994}. For shower energy at collider energies the
situation is essentially the one the time-of-flight PET literature faced for
\emph{timing} a decade ago, until Cram\'er--Rao analyses of photon-counting
statistics~\cite{Seifert2012} became that field's standard. Shower \emph{energy}
deserves the same treatment, and the question ``how much information is in a
jet?''~\cite{DattaLarkoski2017} has a detector-side answer waiting.

This paper supplies the treatment. Its contributions:
\begin{enumerate}
\item \textbf{The information-theoretic limit of the detector itself}
(Sec.~\ref{sec:framework}): the model-conditional record floor $\floor{E;\rec}$,
the best achievable energy variance given a specified
record $\rec$ of the cascade, stated in two rigorously distinct but bridged forms. The
first is a local Cram\'er--Rao lower bound, prior-free, which no single estimator of a
summed record attains uniformly in $E$ unless the missing energy is Gaussian. The
second is the operational Bayes form under an explicit prior, the achievable floor,
which is indeed what the reconstruction literature measures. ``Model-conditional''
is the name of the game in reconstruction and we formalize the factorization of
software methods from hardware capabilities. Our results are conditional on the cascade model and
the record specification, as they are in every compensation and limits study that has preceded them.
\item \textbf{The structure of the detector record} (Sec.~\ref{sec:theorems}): material
interactions, which we refer to as records, form a lattice under refinement on which
information is monotone. Under this lattice structure, independent thinning of a Poisson record
(sampling fraction, photodetection) scales Fisher information exactly by the retention,
clustered records read out as sums cost asymptotically the factor $1/(1+\mathrm{cv}^2)$
underlying the classic sampling-fluctuation term, and chained additive degradations obey a
two-sided bound whose Gaussian-equality case is the field's universal quadrature
budget. Loss channels statistically independent of the record admit a hard lower bound
that survives, in weakened form, the physical mixed case in which many events carry no
such loss.
\item \textbf{Robust estimators} (Sec.~\ref{sec:hadronic},
App.~\ref{app:protocol}): floors are never quoted from a single fit. Every quoted
resolution is an achieved value and bounds its record's floor from above, produced by
estimators whose ability to reach exact limits is measured on exactly solvable cascade
models, including a control test in which the estimator is not handed sufficient
statistics. We treat the neutrino floor is an estimate, not a bound on the fixed-energy rms, i.e.
the untaggable-channel posterior width on the cascade models, a truth-assisted residual on
Geant4, labeled as an estimate wherever it appears.
\item \textbf{The energy resolution limit} (Secs.~\ref{sec:em}--\ref{sec:hadronic}):
measurements of the
achieved resolution on Geant4 truth-level records of hadronic showers at collider
energies, from the dissipated-energy sum, through the dual-readout functionals, to
time-, position- and multiplicity-resolved records, with the marginal information
exchange rate of each readout dimension (gate length, granularity, light yield) and an
attribution of the full-record residual by accounting channel. The electromagnetic case is
the contrast: there the quanta-counting limit sits orders of magnitude below
instrumental reality, and the record-level value within a small factor of it.
\item \textbf{Design consequences} (Sec.~\ref{sec:design}): a dimensionless information
efficiency $\eta$ that separates hardware ceilings from algorithmic gaps, which we believe is
the number detector-optimization programs~\cite{Chung2026Agentic} should report against, and an
empirical ranking of candidate next readout functionals with the relevant conditions.
\end{enumerate}

We take the opportunity to make two statements before the study is described in detail.
First, there is no universal scalar ``detector information''. Floors are conditional
on the estimand, the record, and the cascade model, and we state the conditions
correspondingly. Second, the aim of this study is to highlight the importance of reconstruction
alongside hardware capabilities by making the limit explicit. Algorithms are how real
detectors approach their record's floor, and the point is to know where the floor is,
so that hardware R\&D can be aimed at moving it, which is the only thing hardware
can do that software cannot.

\section{Records, ledgers, and two types of limit}\label{sec:framework}

\subsection{The dissipation record}
Fix an absorber configuration and a cascade model. A shower is a realization of a
marked point process
\begin{equation}
  \Dis \;=\; \{(x_i, t_i, \varepsilon_i, m_i)\}_{i=1}^{N},
\end{equation}
where $x_i, t_i$ are the position and time of an elementary dissipation event,
$\varepsilon_i$ its energy, and $m_i$ a locally measurable mark (the ionizing
particle's species class and velocity, hence Birks quenching weight and Cherenkov
eligibility, and not its ancestry). A record specification
\begin{equation}
  \rec \;=\; \rec(\mathcal{C},\, V,\, T,\, \Delta x,\, \Delta t,\, \text{marks})
\end{equation}
declares which dissipation channels $\mathcal{C}$ a technology class can transduce,
over what volume $V$, within what gate $T$, at what spatial and temporal resolution,
retaining which marks. A real instrument is a stochastic kernel applied to the record,
through thinning (sampling fraction, photodetection efficiency), blurring (granularity,
integration), noise, and quantization.

Record specifications form a lattice under refinement, and information about the
primary particle is monotone along it (Sec.~\ref{sec:theorems}). This single fact
organizes every ``dimension of information''. Gate length, cell size, light yield, and
added channels are moves on one lattice, and the marginal information exchange
rates $\partial \Ifish/\partial \log T$, $\partial \Ifish/\partial \log \Delta x$,
$\partial \Ifish/\partial \log N_{\mathrm{pe}}$ are well-defined, measurable, and
mutually comparable -- they are the quantitative form of the concept of ``physical information available
per unit volume''.

\subsection{Record-intrinsic and instrument limits}
We assert two types of limit that must not be conflated: the \emph{record-intrinsic} limit is set by the
full record $\Dis$ available to a technology class and is a property of absorber
physics alone, and no instrument of that class can beat it. The \emph{instrument} limit
applies to an actual readout $X = K(\Dis)$. Their ratio,
\begin{equation}
  \eta \;=\; \frac{\floor{E;\Dis}}{\floor{E;X}} \;\in\; (0,1],
\end{equation}
is the information efficiency of a design, the detective quantum efficiency of
cascaded-systems analysis~\cite{Rabbani1987} transposed to energy estimation. It
factorizes along the readout chain, $\eta = \eta_1\,\eta_{2|1}$ for
$X = K_2(K_1(\Dis))$ with each stage's efficiency measured against its predecessor's
record, and it separates what hardware forgoes (instrument vs. record-intrinsic) from
what algorithms leave unextracted (achieved vs. instrument).

\subsection{The information floor}
Let $M \equiv \EB + \Eesc + \Enu$ denote the missing ledger and suppose the record
determines the visible energy $V$ (the general case only adds notation). The floor has
two formulations which we denote as local and operational.

\emph{Local (prior-free).} At fixed $E$, for estimators $\hat E = V + g(\rec)$ that are
locally unbiased, $\mathrm{Var}(\hat E) \ge 1/\Ifish(E;\rec)$ with $\Ifish$ the Fisher
information of the record family. Because $V = E - M$, the naive-record family is a
location family when the missing-energy fluctuation is locally $E$-independent.
Writing $M = \alpha E + \epsilon$ with $\alpha$ the mean missing fraction and
$\epsilon$ its fluctuation, $\Ifish(E; \{V\}) = (1-\alpha)^2 J(f_\epsilon)$, i.e. the
response slope times the Fisher information of the fluctuation density (which is the relevant regime
in this study, though not at the lowest energies, where $\alpha$ itself drifts over a resolution
width). This is a lower bound only. For the whole-line location model, with
$E \in \mathbb{R}$ and a fixed centered fluctuation law, the minimax risk of a summed
record, one observation per event, is $\sigma_\epsilon^2/(1-\alpha)^2$, attained by the
affine calibration $\hat E = V/(1-\alpha)$. The Bayes risk converges to the same value
along a locally flattening prior sequence with suitable tail control, and the affine
rule, the posterior mean under the improper flat prior, is asymptotically Bayes along
such sequences. The Cram\'er--Rao value is attainable uniformly in $E$ only for
Gaussian fluctuations. Point-tailored locally unbiased estimators attain it at a chosen
energy and at the cost of performance elsewhere. The operational floor of any summed readout is therefore the
variance of the missing energy, in the whole-line minimax and diffuse-prior senses,
whatever the shape of its distribution. Heavier-than-Gaussian tails in $M$ raise
$J(f_\epsilon)$ above $1/\sigma_\epsilon^2$, but that information cannot be collected
from the sum alone. It becomes collectable only through records that resolve the
components of $M$, which is what the richer records below do, which tilt away from pure location.

\emph{Operational (what experiments quote).} Under an explicit prior $\pi(E)$, which we take
as log-uniform over the study range, the Bayes floor is $\MMSE(E \mid \rec)$, and the
measurement procedure is that of the current reconstruction literature: Train an estimator across
the spectrum, evaluate residuals on fixed-$E$ ensembles in the prior interior, and check
bias. Ledger duality holds in both forms. Estimating $E$ and predicting $M$ are the
same problem, so the floor decomposes by ledger channel and by record content.

The two formulations are bridged by the so-called van Trees inequality. The prior-integrated
Bayes risk is bounded below by $1/(\mathbb{E}_\pi[\Ifish] + \Ifish(\pi))$, with
$\Ifish(\pi)$ the prior's own Fisher information, and equality holds when the posterior
is Gaussian with data-independent variance. The target the operational procedure
approaches is therefore the exact Bayes floor, and how closely it approaches is measured
on cascades where the floor is exactly computable
(App.~\ref{app:protocol}). As a notation convention, every floor quoted from data in
this paper is bracketed. The upper side is the resolution a trained estimator actually
achieves, while the lower side is
an exact bound only where the hypotheses of Sec.~\ref{sec:nu} hold, and in the Geant4
study it is an estimated neutrino floor, labeled as an estimate throughout.

\section{Structure theorems}\label{sec:theorems}

Throughout this work, $\Ifish(E;\rec)$ denotes the Fisher information about the incident energy
carried by record $\rec$. Proofs are collected in App.~\ref{app:proofs}, and every
claim in this section is additionally verified numerically on generative models where
exact answers are computable (App.~\ref{app:protocol}).

\subsection{Monotonicity on the record lattice}
\begin{prop}[Refinement monotonicity]\label{prop:dpi}
If $\rec'$ is obtained from $\rec$ by any stochastic kernel not depending on $E$
(coarsening, thinning, noise, quantization, functional readout), then
$\Ifish(E;\rec') \le \Ifish(E;\rec)$, and likewise for mutual information.
\end{prop}
This is the data-processing inequality in its Fisher form~\cite{Zamir1998}, and its
consequences do the organizing work that follows. The instrument limit cannot beat the
record-intrinsic limit for any readout. Information is monotone in gate length, cell
fineness, light yield, and channel additions for nested records (ones that retain the
finer gates or cells alongside the coarser). The marginal exchange rates of
Sec.~\ref{sec:framework} are therefore well-defined objects of measurement, and ``new
hardware moves the bound, algorithms approach it'' is a theorem.

\subsection{Thinning is exact}
\begin{prop}[Retention scaling]\label{prop:thin}
Let the record be an observed Poisson point process with intensity $\lambda_E$ on mark
space $\mathcal{S}$, and let each point be independently retained with known,
$E$-independent probability $p$. Then
$\Ifish_{\mathrm{thin}}(E) = p\,\Ifish_{\mathrm{full}}(E)$, exactly, for any $p$.
\end{prop}
The proof is one line from the Poisson likelihood.
$\Ifish(E) = \int (\partial_E \lambda_E)^2/\lambda_E$, and thinning maps
$\lambda_E \to p\lambda_E$. In the dilute limit where deposited quanta are
uncorrelated, sampling fraction equals information fraction, and the hypotheses
carry the content. Clustering (below), occupancy, $E$-dependent efficiency, and analog
summation each break exactness by computable amounts. The scaling is confirmed
numerically to better than $0.2\%$ for $p \in \{1, 0.3, 0.02\}$.

\subsection{Clustered records and effective quanta}\label{sec:cluster}
Real records are clustered. One track crossing an active layer deposits a correlated
cluster of quanta. Let the cluster count be $N \sim \mathrm{Poisson}(\mu(E))$ with
i.i.d.\ cluster masses $M$ whose law does not depend on $E$ and is absolutely
continuous, with a density of finite Fisher information and finite moments, and let the
readout observe only the sum $S = \sum M_i$.
\begin{prop}[Clustering cost, normal regime]\label{prop:cluster}
As $\mu \to \infty$,
$\Ifish_S(\mu) = \dfrac{1}{\mu}\dfrac{1}{1+\mathrm{cv}^2}\big(1 + o(1)\big)$,
where $\mathrm{cv}^2 = \mathrm{Var}(M)/\langle M\rangle^2$, while the cluster-resolved
record, which includes the count $N$, gives $\Ifish = 1/\mu$ exactly.
\end{prop}
The smoothness hypothesis carries the content. For masses taking two positive values in
an irrational ratio the sum identifies the count and $\Ifish_S = 1/\mu$ for every
$\mathrm{cv}^2$, so no moment condition alone delivers the asymptotic. The effective
number of quanta is $N/(1+\mathrm{cv}^2)$, the classic sampling-fluctuation
phenomenology as an information statement. Exact evaluation via the characteristic
function confirms the asymptotics for unit-mean gamma masses. At $\mu = 2000$,
$\Ifish_S\,\mu$ reaches $0.8000$, $0.5001$ and $0.2002$ for
$\mathrm{cv}^2 = 0.25$, $1$ and $4$. The finite-$\mu$ corrections are shape-dependent
and several times smaller than the Gaussian-surrogate term of App.~\ref{app:proofs},
and the finite-information hypothesis is sufficient but not necessary, since the gamma
laws at $\mathrm{cv}^2 = 1$ and $4$ have infinite location information yet reach the
asymptotic. The same evaluation quantifies the shape corrections. For a heavy-tailed
lognormal mass distribution with $\mathrm{cv}^2=3.2$ the exact sum information
exceeds the two-moment normal-regime prediction by $6\%$ at $\mu=200$.
Distributional shape carries information that a quadrature budget ignores, while
remaining below the resolved-record value, which is what cluster-resolving readout
(granularity, timing) recovers, statistically.

\subsection{The composition calculus and the quadrature budget}
\begin{prop}[Two-sided budget]\label{prop:stam}
For a chain of independent additive location degradations
$X = E + \sum_k \epsilon_k$ with densities $f_k$ of finite Fisher information
$J_k$ and variance $\sigma_k^2$, observed only at the chain's end,
\[
  \sum_k J_k^{-1} \;\le\; J\!\big(f_1 * \cdots * f_n\big)^{-1} \;=\; \CRB_{\rm chain}
  \;\le\; \sum_k \sigma_k^2 ,
\]
with all three equal iff every stage is Gaussian.
\end{prop}
The left inequality is Stam superadditivity~\cite{Stam1959}. The right is
$J\sigma^2 \ge 1$ per stage applied to the convolution. Read at fixed $E$ with the
stage scales frozen (an energy-dependent noise scale $v(E)$ adds a $v'^2/2v^2$ term to
the information that this location model does not carry), the ubiquitous budget
$a/\sqrt{E} \oplus b \oplus c/E$ is thereby the Gaussian-equality case. Plain variance
quadrature needs no Gaussianity, but its saturation of the information bound
does. The headroom between $\sum\sigma_k^2$ and $\CRB_{\rm chain}$ that non-Gaussian
stages open cannot be collected from the summed observation itself
(Sec.~\ref{sec:framework}: at one observation per event the achievable floor of a
summed record is its variance). It becomes collectable only through records that
resolve the stages before summation, and the lattice of Sec.~\ref{sec:framework} is the
bookkeeping of what each such record contributes. The familiar error budget thereby acquires
an exact status. It is the achievable floor of every summed readout and the upper side
of a two-sided information bound, and the two coincide exactly when every fluctuation
is Gaussian.

\subsection{Untaggable channels: hard floors}\label{sec:nu}
\begin{lem}[Location-smoothing bound]\label{lem:nu}
Suppose the experiment factors as the Markov chain $E \to E - \nu \to \rec$ with kernel
independent of $E$, $\nu$ independent of the cascade variables entering $\rec$, and
$f_\nu$ absolutely continuous with finite location-Fisher information $J(f_\nu)$. Then
for every such record, $\Ifish(E;\rec) \le J(f_\nu)$, i.e.\
$\CRB(E;\rec) \ge 1/J(f_\nu)$.
\end{lem}
Physical loss distributions are mixed in reality. With probability $q_0$ no loss occurs at
all (no decay in flight), so $J(f_\nu) = \infty$ and the bound is vacuous, correctly,
since the no-loss class is in principle recoverable. Conditioning on the loss count $D$
(which can only lower the MMSE) gives the operative mixed-measure bound
\[
  \MMSE(E \mid \rec) \;\ge\; \sum_{d \ge 1} \frac{P(D{=}d)}{J(f_{\nu|d}) + \Ifish(\pi)},
\]
with each $f_{\nu|d}$ absolutely continuous and $\Ifish(\pi)$ the prior's own Fisher
information (App.~\ref{app:proofs}; negligible for broad priors). In terms of detetor design, a
record feature that tags loss occurrence (a muon stub,
a decay kink) removes this channel's penalty from the $d{=}0$ class, whose energy is
then limited only by the rest of the record, and pays only the $d{\ge}1$ penalty. For
real cascades neutrino emission is not exactly independent of the record, since the
muon from $\pi \to \mu\nu$ is in it, so in the measurements below the untaggable
channel enters as an estimated irreducible component, with the Lemma reserved for
channels satisfying its hypotheses.

\section{Electromagnetic showers: an instrument-limited regime}\label{sec:em}

For electromagnetic showers the ledger is almost empty. There are no nuclear binding
losses at leading order and no neutrinos, and containment is an engineering choice. The
record-level value therefore approaches the quanta-counting (Fano) limit far more
closely than the hadronic one does (for BGO at 50 GeV the two gaps are 1.5 and 2.2
decades). With $W_{\rm eff}$ the mean deposited energy per transducible quantum and $F$
the Fano factor, $\sigma/E|_{\rm quanta} = \sqrt{F\,W_{\rm eff}/E}$.
Table~\ref{tab:fano} collects representative values. For fully active media the
quanta-counting limit lies at the $10^{-6}$--$10^{-4}$ level at collider energies, two
to three orders of magnitude below the achieved constant terms of $0.3$--$0.5\%$ of the
best crystal calorimeters.

\begin{table}[b]
\caption{Quanta-counting floors $\sqrt{F W_{\rm eff}/E}$ for representative fully
active media. $W_{\rm eff}$ is per detected-class quantum (ionization pairs or
scintillation photons). $F$ is the Fano factor for ionization-pair counting; for
scintillation-photon counting the corresponding relative-variance factor is
$\approx 1$ (near-Poisson), the value the BGO/PWO rows use; scintillator
non-proportionality, not counted here, lifts the practical scintillation limit.
Inputs: $W$ for Si and LAr and the relative scintillation light yields from
Ref.~\cite{PDG2024}; the Fano factors are the standard measured values for
ionization in Si and LAr ($F \approx 0.11$--$0.12$); for the scintillators
$W_{\rm eff} = 1/(\text{absolute light yield})$ with the relative yields scaled to
NaI(Tl) at $38$ photons/keV (BGO $\approx 21\%$, PWO $\approx 0.3\%$), hence the tildes.}
\label{tab:fano}
\begin{ruledtabular}
\begin{tabular}{lccc}
medium & $W_{\rm eff}$ & $F$ & $\sigma/E$ at 100 GeV \\
\hline
Si (e--h)        & 3.65 eV  & 0.12 & $2\times10^{-6}$ \\
LAr (ion pairs)  & 23.6 eV  & 0.11 & $5\times10^{-6}$ \\
BGO (photons)    & $\sim$120 eV & $\sim$1 & $3\times10^{-5}$ \\
PWO (photons)    & $\sim$10 keV & $\sim$1 & $3\times10^{-4}$ \\
\end{tabular}
\end{ruledtabular}
\end{table}

The quanta-counting limit is separated from the record floor by the small stochastic
traffic of the electromagnetic cascade into nuclear channels, photonuclear interactions
with their ejected neutrons and residual binding losses, plus albedo and sub-threshold
dissipation, which we measure in simulation. Electron showers in copper spend on average
$0.30\%$ of their energy on photonuclear binding at 50 GeV ($0.23\%$ at 10 GeV,
App.~\ref{app:g4}), and the achieved values at 50 GeV, obtained with the same
trained-estimator procedure on $e^-$ records and bounding the electromagnetic record
floor from above, are $0.12\%$ rms ($0.08\%$ Gaussian core) in Cu and $0.14\%$
($0.09\%$) in BGO, with unresolved photonuclear binding the leading residual channel.
The rms/core split is the heavy photonuclear tail, the non-Gaussian missing-energy
distribution of Sec.~\ref{sec:framework} made empirical. To our knowledge these numbers
have not been previously quoted as electromagnetic record-level values. The full hierarchy
is then: quanta-counting limit $\sim 0.005\%$ (BGO, 50 GeV), achieved record-level
value $\sim 0.1\%$, best achieved crystal constant terms $0.3$--$0.5\%$.
Electromagnetic calorimetry has engineered its way to within a small factor of its
\emph{record}-level value, with the remaining gap at most that factor and instrumental
in origin, while sitting far above its quanta-counting limit. The photonuclear
excursions that set the electromagnetic record floor also contaminate the
\emph{hadronic} one-functional limit construction, as Sec.~\ref{sec:llwrepro} shows.
The remaining gap between the record-level value and achieved performance is instrumental:
photostatistics (a thinning, Prop.~\ref{prop:thin}), light-collection nonuniformity,
intercalibration, and upstream material. We find the design conclusion is structural:
\emph{Electromagnetic calorimetry is instrument-limited}, and its resolution frontier
advances by engineering the readout chain toward $\eta \to 1$, \emph{while hadronic
calorimetry is record-limited}. Within the sensitivity of this analysis, the hadronic
frontier advances by enriching the record itself. Through this framework, the
field's entire division of effort between crystal engineering and hadronic-technique
invention is explained by which side of its record floor each sits on.

\section{Hadronic showers: the ladder to the floor}\label{sec:hadronic}

\subsection{Setup and the per-event energy ledger}
Single $\pi^-$ (with $e^-$ controls) are absorbed in a $1.6\times1.6\times3.2$~m block
of copper ($\approx 10.3\times10.3\times20.5\,\lambda_I$ with $\lambda_I = 15.59$~cm
from the simulation's material table; the PDG tabulates 15.32~cm), simulated with
Geant4 11.4.2~\cite{GEANT4} and FTFP\_BERT with the neutron tracking cut removed. The
default kill at 10~$\mu$s silently discards late-capture information, which is exactly
the record content under study. For every event the complete ledger is recorded:
ionizing and non-ionizing deposits with positions and times, escapes and albedo by
species, per-vertex kinetic-ledger imbalances $Q_v$ (binding), neutron creation,
capture, and escape, and the em-tagged component $\fem$ with full-ancestry
de-duplication. The per-event energy balance,
$E_{\rm ref} = \Edis + \sum \mathrm{esc} + \sum Q$, holds at machine precision (typical
events at $10^{-14}$--$10^{-12}$ relative; production rms $\le 2\times10^{-11}$
within the loader's $10^{-6}$ acceptance, App.~\ref{app:g4}), so a nonzero residual signals unaccounted energy instead of
entering any calibration. Records at each level are computed from the same events:
Birks-weighted $S$ and Cherenkov-eligible $C$ functionals, longitudinal and radial
profiles, six time gates from 10~ns to $\infty$, and capture-$\gamma$ and dense-hit
counts. Training uses a log-uniform 8--120~GeV spectrum, and evaluation uses disjoint
fixed-energy runs, following the validated procedure of App.~\ref{app:protocol}.
Table~\ref{tab:recdef} fixes every record-element definition the record levels are
built from, and the complete set of conventions is documented with the code.

The record elements have different experimental status, and we state the distinction
before quoting any resolution. Dual-readout calorimeters measure scintillation
and Cherenkov signals~\cite{WigmansRMP2018,CalVision2024}, granular calorimeters
measure shower profiles~\cite{CALICE2012SC,CALICE2013G4Val}, and the neutron-sensitive
time structure of hadronic showers has been
measured~\cite{Akchurin2009Neutron,CALICET3B2014}. The $S$, $C$, profile, and gated
elements here are idealized versions of those signals, and the record additionally
retains the exact unquenched ionization sum $\Sigma_\infty$, which a light-only
$\{S, C\}$ record lacks. Both idealizations are analyzed in Sec.~\ref{sec:exchange}
(bins only against exact moments; light only against anchored). Two elements have no
instrumental counterpart. The per-gate counts of energy-depositing simulation steps in
$\mathcal{R}_3$, and the dense-step counts and creation-tagged capture $\gamma$s of
$\mathcal{R}_4$, stand in for detector hit multiplicity and reconstructed capture tags,
and that correspondence is not modeled. The count-free record is therefore the one to
read for hardware: $\mathcal{R}_2$ reaches $1.60\%$ at 20 GeV ($1.61\%$ with native
bins only) against $1.29\%$ for the full record, and removing the $\mathcal{R}_4$
additions moves the full-record value to $1.32\%$, within the cross-run spread
(Table~\ref{tab:ladder}). Binding losses, escaped energy, and neutrino energy are never
reconstruction inputs. They enter as truth-assisted diagnostics, labeled as such. The
physics-list comparisons of Sec.~\ref{sec:outlook} quantify the model sensitivity of
these values (under retrained high-precision neutron transport the full record moves by
$0.4\%$ and the other record levels by up to $9\%$, relative), and whether the
correlations the estimators exploit are those of real showers is a question for
test-beam data, not for a second simulation. These are isolated single-particle results
without pileup, the regime in which the question is cleanly posed. Jets, pileup, and
particle flow require different records and estimands (Sec.~\ref{sec:outlook}).

\begin{table}[t]
\caption{Record-element definitions (conventions, abridged). Deposit time
is post-step global time; $\lambda_I(\mathrm{Cu}) = 15.59$~cm; production range
cut 0.7~mm; MIP $= 1.403~\mathrm{MeV\,cm^2/g}$ (mass stopping power)
$\times\,\rho$. The record levels of
Table~\ref{tab:ladder} are cumulative: $\mathcal{R}_0 = \{\Sigma_\infty\}$;
$\mathcal{R}_1$ adds $\{S_\infty, C_\infty\}$; $\mathcal{R}_2$ adds profiles and
moments; $\mathcal{R}_3$ adds the five finite-gate sums and hit counts plus the
gated longitudinal profile (the $\infty$-gate hit count is not included);
$\mathcal{R}_4$ adds capture-$\gamma$ and dense-hit
counts. The last bin of each profile doubles as overflow. These conventions are
part of the record specification itself, not nuisance parameters, and the
optional step-level event dump (\texttt{/ledger/dumpSteps}) permits re-derivation
under alternatives by rerun from the recorded macros and seeds (the archived
productions store event-level records).}
\label{tab:recdef}
\begin{ruledtabular}
\begin{tabular}{ll}
element & definition \\
\hline
$S$ & $\sum_{\rm charged} e_{\rm ion}/(1 + k_B'\,\mathrm{d}E/\mathrm{d}x)$; \\
    & \enskip $k_B = 0.0126~\mathrm{g/(MeV\,cm^2)}$, $k_B' = k_B/\rho$; \\
    & \enskip neutral-step deposits unquenched; \\
    & \enskip at-rest deposits fully quenched \\
$C$ & $\sum \ell_{\rm step}\, z^2$, charged steps \\
    & \enskip with $\beta > 1/n$, $n = 1.47$ \\
gates & $\{10, 50, 100, 500~\mathrm{ns},\ 10~\mu\mathrm{s},\ \infty\}$ \\
long.\ profile & $42 \times 0.5\,\lambda_I$ bins ($\infty$ and 100~ns gates) \\
radial profile & $24 \times 0.25\,\lambda_I$ bins ($\infty$ gate) \\
moments & $\langle z \rangle$, $\mathrm{rms}\,z$, $\langle r \rangle$, $\mathrm{rms}\,r$, \\
    & \enskip each weighted by $e_{\rm ion}$, $S$, $C$ \\
capture-$\gamma$ & neutron-capture $\gamma$s at creation, per gate \\
hit counts & energy-depositing steps per gate; \\
    & \enskip dense: steps with $\mathrm{d}E/\mathrm{d}x > 5 \times \mathrm{MIP}$ \\
\end{tabular}
\end{ruledtabular}
\end{table}

\subsection{Reproducing and dissecting the canonical limit}\label{sec:llwrepro}
Before extending Ref.~\cite{LeeLivanWigmans2018}, we reproduce it. At 20~GeV in copper
the construction of Ref.~\cite{LeeLivanWigmans2018}, the event-by-event spread of
nuclear binding loss around its one-functional proxy, gives $\mathrm{rms/mean} = 0.162$
for $\Delta B / E_{\rm non\text{-}em}$. Our production (the 12k-event 20 GeV sample)
yields $0.153$ once one structural subtlety is taken into account. The literal
ratio in our accounting has $\mathrm{rms/mean} = 0.43$
($0.71$ in an early 2k-event diagnostic; the tail is sample-hungry), driven by a small
population in which the numerator's energy source sits inside the
electromagnetic component. Hard photonuclear interactions of shower photons
re-inject GeV-scale hadronic cascades, whose binding losses
($\langle 80 \rangle$~MeV at 20~GeV, with excursions to $3.7$~GeV in the 12k events)
belong to the em-tagged descent. Splitting the binding ledger by ancestry,
$\Delta B = \Delta B_{\rm had} + \Delta B_{\rm em}$, brings the hadronic-side
conditional width to within about $5\%$ of the canonical value ($0.153$ against the
$0.162$ of Ref.~\cite{LeeLivanWigmans2018}). We take the rms/mean of
$\Delta B/E_{\rm non\text{-}em}$, the inverse of the ratio displayed in Fig.~7 of
Ref.~\cite{LeeLivanWigmans2018}. Inversion is sensitive to near-zero binding losses and
need not preserve the relative width, and in a controlled small-fluctuation expansion
its leading relative correction scales with the skewness times the cv. The split also
exposes that the one-functional construction is blind to the em-side invisible energy
by design, while the record is not: photonuclear events leave neutron and
capture-$\gamma$ signatures. The ``ultimate limit'' of Ref.~\cite{LeeLivanWigmans2018}
is thereby reproduced and interpreted as a first rung on a ladder. It conditions on one variable, reads
one side of the ledger, and rests on a linearized error propagation that fails on
charge-exchange-like events with vanishing non-em energy. None of this diminishes the
physics insight of that analysis. It is the reason a floor must be defined at the record level and
measured with estimators, as follows.

The construction evaluated on this study's events is also the right anchor for factor
comparisons, because this study's $2.14\%$ and the published $2.7\%$ at 20~GeV differ by
bookkeeping and model era, not by physics disagreement. The construction evaluates to
$(\mathrm{rms/mean}) \times \langle \Delta B/E \rangle$. This study reads
$0.1531 \times 0.140 = 2.14\%$, against $0.162 \times 0.164 = 2.7\%$ for the example
worked in Sec.~5 of Ref.~\cite{LeeLivanWigmans2018} following its Figs.~8 and 9
(Geant4 10.3p02, FTFP\_BERT, $10^4$ $\pi^-$ per energy on a massive monoisotopic
$^{63}$Cu block; ours, Geant4 11.4.2 on natural copper). Three-quarters of the
logarithmic gap is the mean binding fraction ($0.164 \to 0.140$): nine years of
hadronic-model evolution (Geant4 10.3p02, 2017, to 11.4.2, 2026), the isotope
composition, and the ancestry split, which removes
$\langle \Delta B_{\rm em} \rangle \approx 80$~MeV at 20 GeV from the hadronic
numerator. Ref.~\cite{LeeLivanWigmans2018} describes its $f_{\rm em}$ as a creation sum
over $\pi^0$s and $\gamma$s, and does not specify whether descendants of photonuclear
re-injection are de-duplicated, so the ancestry-split quantity is a modified same-events
benchmark rather than a reconstruction of that numerator. The remainder of the gap is
the conditional width ($0.162 \to 0.153$). In an accounting with complete photonuclear
ancestry the \emph{unsplit} construction is not usable at all, since its ratio
distribution acquires the divergent tail above, so the split is what makes the
comparison well-posed. The anchor is also generator-stable. Recomputed on QGSP\_BERT
and FTFP\_BERT\_HP samples at 20 GeV it reads $2.16\%$ and $2.11\%$ ($2.08\%$ on the
clean high-precision configuration of Sec.~\ref{sec:outlook}) against the baseline
$2.14\%$, shifts of $3\%$ or less, with the em-descended binding population equally
stable ($\langle \Delta B_{\rm em} \rangle = 78$--$89$ MeV across the three lists).
Every factor claim in this paper is therefore anchored to the value from the study's own
events, with the published $2.7\%$ quoted for context.

\subsection{The ladder}\label{sec:ladder}

\begin{figure}[t]
\includegraphics[width=\columnwidth]{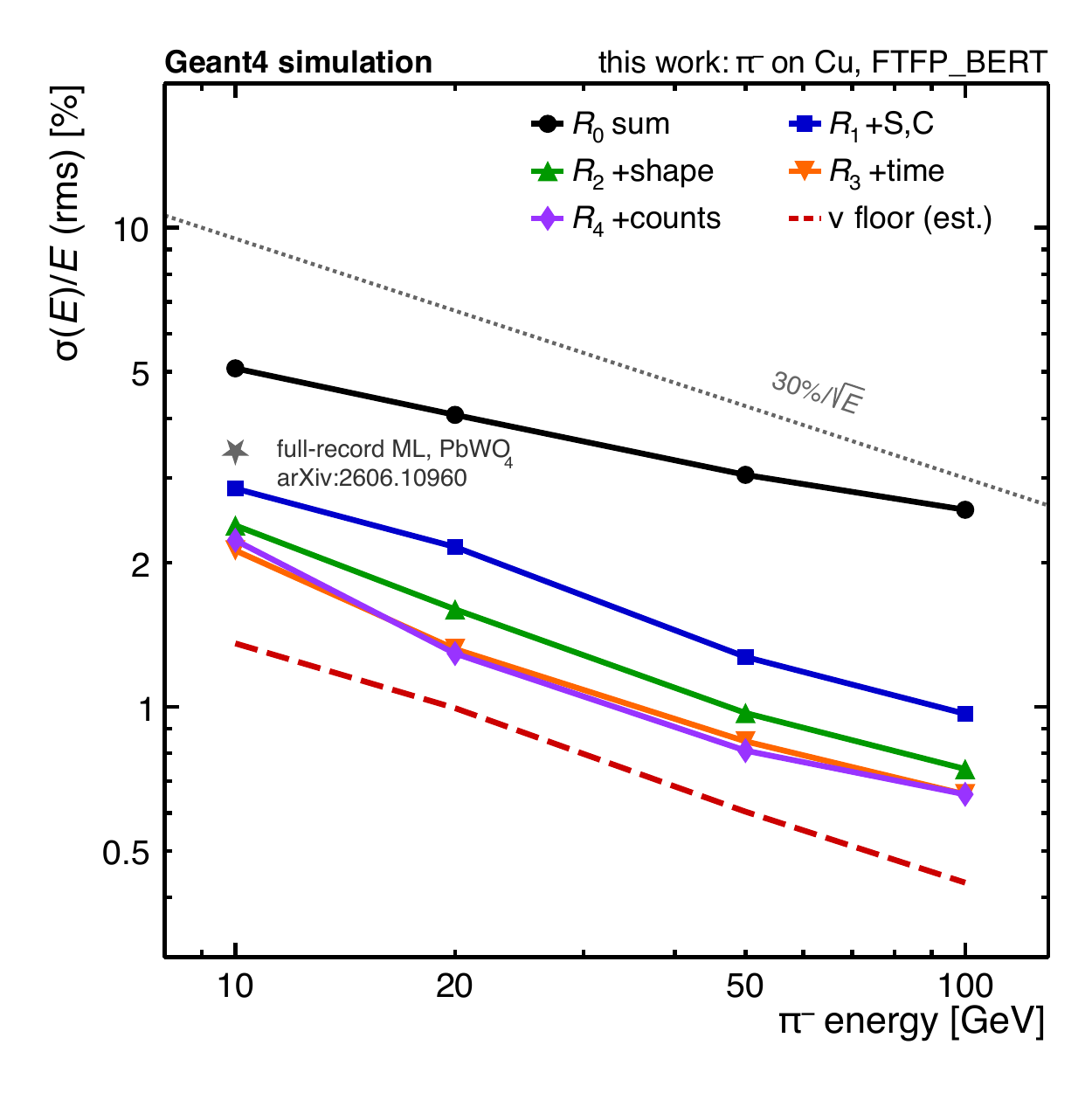}
\caption{The achieved-resolution ladder: $\sigma/E$ (best of gradient-boosted trees
and multilayer perceptron, rms) versus $\pi^-$ energy in copper for the record levels
$\mathcal{R}_0$--$\mathcal{R}_4$. Each point is an achieved value and bounds its
record's floor from above. The dashed curve is the estimated neutrino floor (an
estimate, not a derived bound), and the $\le$10 GeV full-record ML point of
Ref.~\cite{Fudan2026ML} is shown for context (different absorber and record
definition; not directly comparable). Points are the best of two independent training
runs (text).}
\label{fig:ladder}
\end{figure}

\begingroup
\squeezetable
\begin{table}[b]
\caption{Achieved resolution $\sigma/E$ [\%] per record level (rms, with
Gaussian-core robust width in parentheses), $\pi^-$ in Cu, feature families as in
Sec.~\ref{sec:hadronic}, procedure as in App.~\ref{app:protocol}. Each cell is the
best of two independent training runs (log-uniform 8--120 and 5--130~GeV, 200k events
each), evaluated on disjoint 12k-event fixed-energy runs; the robust width is half the
central-68.27\% interval (15.87th to 84.13th percentiles). Cross-run rms spreads at
20~GeV are 0.04, 0.03, 0.07, 0.09 and 0.19\% for
$\mathcal{R}_0$--$\mathcal{R}_4$; $\approx 0.03\%$ at 50--100 GeV; up to $0.4\%$ at
10 GeV on the richest records (all in absolute $\sigma/E$). App.~\ref{app:runs}
tabulates both runs; spreads are computed before rounding, and since the runs differ
in seed \emph{and} span, the spread bounds the training lottery and prior-support
effects together. The $\nu$ row is an estimated floor (truth-assisted residual of the
full in-volume neutrino ledger, escape plus albedo, given the full record; cross-run
spread $\le 0.04$), not a derived bound: estimator slack can only inflate it. Digits
beyond the spread are carried for comparability, not significance. Resolutions of the
trained estimators here and throughout are normalized to the beam kinetic energy (the
gun label $E$); the one-functional construction of Sec.~\ref{sec:llwrepro} is
normalized to the ledger total $E + m_\pi$ ($2.14\%$ at 20 GeV; $2.16\%$ if
beam-normalized); normalizing the ladder to the ledger total would lower every entry
by a common $0.1$--$1.4\%$ relative ($0.7\%$ at 20 GeV), affecting no comparison.}
\label{tab:ladder}
\begin{ruledtabular}
\begin{tabular}{lcccc}
record & 10 & 20 & 50 & 100~GeV \\
\hline
$\mathcal{R}_0$ sum        & 5.09 (4.96) & 4.07 (3.83) & 3.05 (2.86) & 2.58 (2.42) \\
$\mathcal{R}_1$ +$S,C$     & 2.85 (2.57) & 2.16 (1.79) & 1.27 (1.13) & 0.97 (0.81) \\
$\mathcal{R}_2$ +shape     & 2.39 (2.10) & 1.60 (1.43) & 0.97 (0.89) & 0.74 (0.62) \\
$\mathcal{R}_3$ +time      & 2.12 (1.86) & 1.32 (1.17) & 0.85 (0.75) & 0.66 (0.50) \\
$\mathcal{R}_4$ +counts    & 2.23 (1.83) & 1.29 (1.12) & 0.81 (0.66) & 0.66 (0.51) \\
\hline
$\nu$ floor (est.)         & 1.36 & 0.99 & 0.60 & 0.43 \\
\end{tabular}
\end{ruledtabular}
\end{table}
\endgroup

Table~\ref{tab:ladder} and Fig.~\ref{fig:ladder} are the paper's central measurement,
and three statements summarize them. First, the canonical limit is a rung, far
from the top. The one-functional construction of Ref.~\cite{LeeLivanWigmans2018},
evaluated on this study's own events with ancestry-split binding (Sec.~\ref{sec:llwrepro}),
gives $\sigma_{1f}/E = 2.89$, $2.14$, $1.51$ and $1.21\%$ at 10, 20, 50 and 100 GeV.
The anchored dual-readout level $\mathcal{R}_1$ tracks this construction at 10--20 GeV
($2.85$ vs. $2.89$; $2.16$ vs. $2.14$) and beats it above. Comparisons to light-only
fiber or crystal dual-readout technology should instead use the $\{S,C\}$-only value of
Sec.~\ref{sec:exchange}, $2.78\%$ at 20 GeV. The full record sits a factor
$1.7$--$1.9$ \emph{below} the construction across 20--100 GeV ($1.29\%$ vs. $2.14\%$ at
20 GeV; $0.66\%$ vs. $1.21\%$ at 100 GeV; with single-run values in place of the
best-of, the range is $1.45$--$1.9$), approximately
$\sigma/E \simeq 5.5\%/\sqrt{E} \oplus 0.3\%$ over 20--100 GeV. An unweighted
two-parameter fit to the three rms points reads $5.5\%/\sqrt{E} \oplus 0.31\%$, and the
Gaussian core prefers $4.9\%/\sqrt{E}$ with a vanishing constant term over this range.
These forms are descriptions to guide the eye, the underlying values being the
table's, and the 10 GeV points sit above them, from growing binding and neutrino
fractions at low energy. Second, the information is unevenly distributed. The
dual-readout pair does the largest single share of the work, halving $\sigma$ relative
to the sum. The next-largest step is the shower-shape family ($2.16 \to 1.60\%$ at
20 GeV, all of it longitudinal, none radial; Sec.~\ref{sec:exchange}), then time
structure ($\to 1.32\%$), with counting marks adding little beyond time. The gate scan
below measures the cost of integration, losing late energy outright, worth up to
a third on the dual-readout record. The ladder increment measures the value of
structure, subdividing retained energy in time. The two record dimensions turn
out nearly additive in achieved $\sigma$ rather than degenerate. Adding the full gated
record to dual-readout without any profile information reaches $1.90\%$ at
20 GeV, and time then improves $\sigma/E$ by $0.26\%$ on its own against $0.28\%$ on
top of shower shape ($1.60 \to 1.32\%$), while shower shape contributes
$0.56$--$0.58\%$ whether or not time is present (2.86, 1.90, 1.13 and 0.89\% across
the four energies for the shape-free gated record). At 10 GeV the time increment is
unresolved within the $0.2\%$ cross-run spreads. At 20--100 GeV neither dimension's
$\sigma$ improvement is diminished by the other's presence, additivity on the $\sigma$
scale at this operating point, and on an information scale ($1/\sigma^2$) the joint
gain compounds, as it must. The $0.99\%$ below the ladder is an estimated floor, and
the gap between it and the achieved $1.29\%$ is not demonstrated headroom. Third,
achieved ladders need not be monotone even though floors are. At 10 GeV the
$\mathcal{R}_4$ rms sits marginally above $\mathcal{R}_3$'s, within the cross-run
spread. We quantify the stability directly by repeating the entire chain with an
independent training run on a different spectrum span. The achieved values reproduce to
within $0.031\%$ at 50--100 GeV and to within $0.4\%$ at 10 GeV on the richest records
(absolute $\sigma/E$), the measured stability of the achieved values themselves (two
estimators can share a common distance to the floor), quoted so that no difference
smaller than it is over-read.

\subsection{Exchange rates and attribution}\label{sec:exchange}

\begin{figure}[t]
\includegraphics[width=\columnwidth]{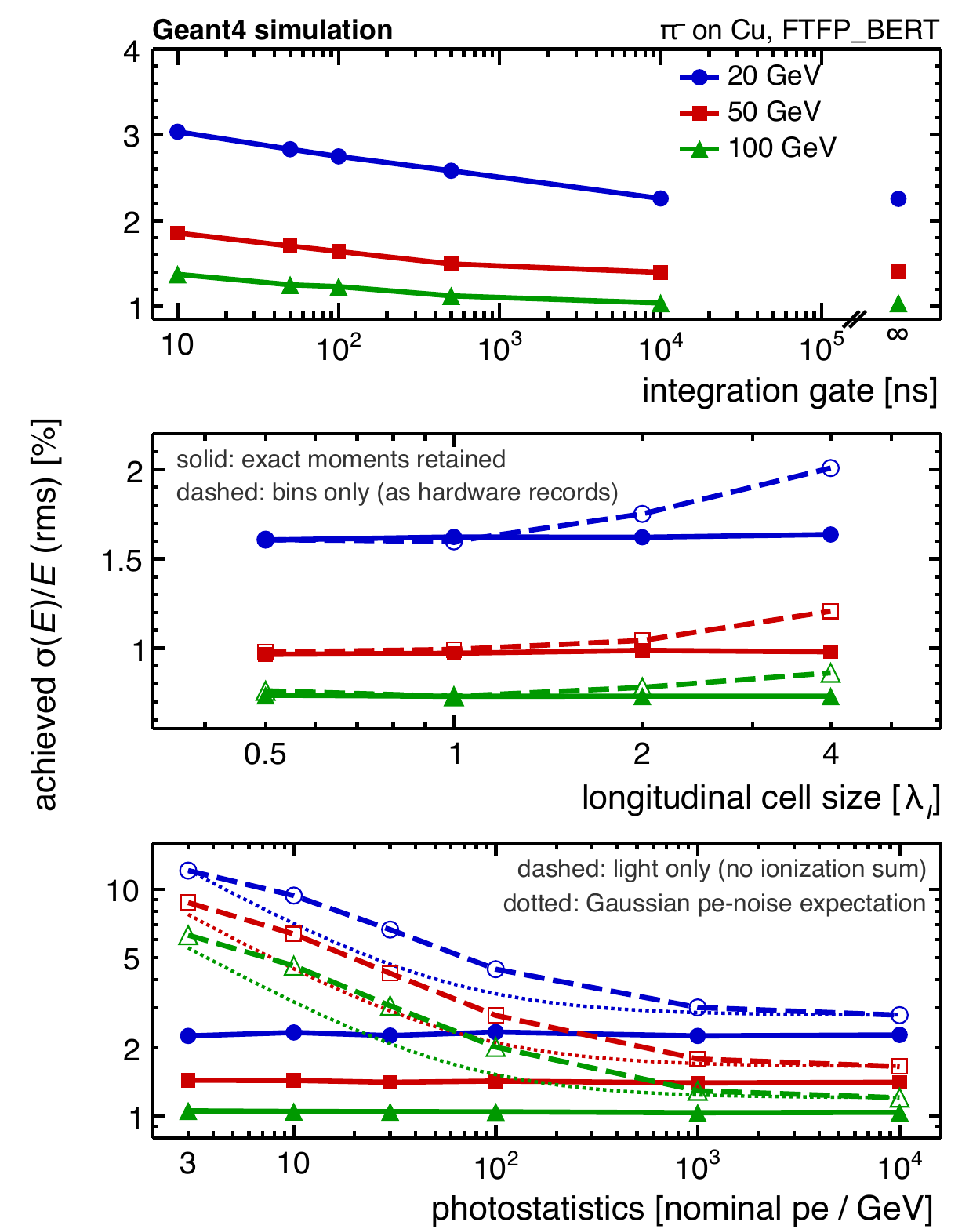}
\caption{Exchange rates of three readout dimensions: integration gate on the
dual-readout record (top; right edge is ungated), profile cell size on the
$\mathcal{R}_2$ record (middle; longitudinal and radial bins coarsened jointly;
solid retains the exact moments, dashed is bins-only, the hardware-realistic
variant), and photostatistics (bottom; log scale; the pe/GeV axis is the nominal
native-unit Poisson convention of Sec.~\ref{sec:exchange}, not a per-channel
optical model): solid, $S,C$
Poisson-thinned with the ionization sum kept; dashed, the light-only record ($S,C$
thinned, no ionization sum, evaluated against its own $S$ baseline); dotted, the
Gaussian pe-noise expectation added in quadrature to each light-only plateau. Cell
sizes are in simulation-$\lambda_I$ units. Values are the best of two independent
training runs; the solid light-yield curve is the minimum over both runs and a
supplementary wide-run scan, which alone provides its 3 and 10 pe/GeV points. The
gate, cell-size and solid light-yield fits are single-family (so the ungated
gate-scan point sits marginally above the best-of $\mathcal{R}_1$); the bins-only and
light-only families select across estimator families. 10 GeV omitted (largest
cross-run spread).}
\label{fig:exchange}
\end{figure}

Figure~\ref{fig:exchange} measures the cost of each ``dimension of information''.
\emph{Time}. A 10~ns gate costs roughly a third in resolution relative to ungated
readout at 20 GeV ($3.04\% \to 2.25\%$ for the tightest values; $32$--$36\%$ within the
individual runs). More than half the recoverable late information arrives by 500~ns and
essentially all of it by 10~$\mu$s. The $\sim$2\% of capture $\gamma$s arriving later
still (App.~\ref{app:g4}) carried no resolving power this analysis could use, and the
ungated and 10-$\mu$s values agree within $0.01\%$ in absolute $\sigma/E$ at
20--100 GeV ($0.22\%$ at 10 GeV). This is the information value of integration time, the quantity behind the
gate-versus-pileup trade every collider design faces.
\emph{Granularity}. The shower-shape information is longitudinal, low-dimensional, and
its sampling scale is now pinned from three directions. Deleting all
longitudinal content from the shower-shape record, profile bins and $z$-moments alike,
costs $38$--$42\%$ at every energy ($1.61 \to 2.25\%$ at 20 GeV, landing back at the
dual-readout level), while deleting all radial content costs $2$--$3\%$ at 20--100 GeV
($4.7\%$ at 10 GeV, inside that cell's cross-run spread of $0.4\%$ in absolute
$\sigma/E$), within or near the spreads.
The shower-shape improvement is longitudinal to the sensitivity of this analysis. How
finely must depth be sampled? With no unbinned quantities retained, bins only, the
realistic model of hardware with $0.5f\,\lambda_I$ cells, the achieved value is flat
from $0.5$ to $1\,\lambda_I$ ($1.61$, $1.60\%$ at 20 GeV, against $1.61\%$ for the
like-for-like full shower-shape record: the twelve exact moments are redundant at
native binning), degrades $\sim 2.5$--$9\%$ at $2\lambda_I$ ($1.75\%$ at 20 GeV) and
$13$--$25\%$ at $4\lambda_I$ ($2.01\%$), relative to native binning, at 20--100 GeV.
The achieved value saturates at a sampling scale of about one interaction
length, roughly twenty depth samples here, and segmentation finer than that, down to
the $0.5\,\lambda_I$ tested, added nothing this analysis could resolve for
single-hadron energy. (A scan that retains the exact moments while rebinning stays
flat to $4\lambda_I$, Fig.~\ref{fig:exchange}, middle, solid, but that measures bins
beyond exact moments, which no coarse detector possesses; the bins-only curve,
dashed, is the hardware statement.) As always this concerns profile readout of a
homogeneous record. Transverse 3D imaging, pattern recognition in jets, and
sampling-noise mitigation are different questions. Neub\"user, Kieseler and
Lujan~\cite{Neubuser2022}, regressing pion energies in a homogeneous lead-glass
calorimeter with convolutional networks on segmented deposits, find that longitudinal
segments of $\le 0.5\,\lambda_I$ and transverse segments of $\le 1.3\,\lambda_I$ are
needed for an optimal energy measurement. The two studies agree that depth sampling at
the half-interaction-length scale suffices. They differ on whether coarser depth
sampling and the transverse dimension cost resolution, and the differences in absorber,
record (truth profiles with exact moments against segmented sums), containment,
estimator family and energy range are each large enough to account for that. This study's scan
does not test finer than its native $0.5\,\lambda_I$ bins, so it cannot exclude gains
below that scale. \emph{Light}. With the ionization sum $\Sigma$ kept unthinned,
Poisson-thinning the $S$ and $C$ functionals from $10^4$ down to $3$ detected pe/GeV
leaves the achieved value unchanged (Fig.~\ref{fig:exchange}, solid). Two conventions
to read the axis by. The emulation draws each functional as an analog Poisson count at
the stated rate per native unit, $N_X \sim \mathrm{Poisson}(y\,X)$ for $X \in \{S, C\}$
in every gate ($S$ in GeV, $C$ in its path-length unit), and feeds $N_X/y$ to the
estimator, an idealization of photostatistics rather than a per-channel optical model,
and a Poisson draw on a random functional rather than the point-process thinning of
Prop.~\ref{prop:thin}. And since $C$ is a path-length functional its effective count
per GeV is far larger than $S$'s, so this scan starves the scintillation channel while
leaving Cherenkov comparatively photon-rich. Per-channel optical budgets are an
instrument-level refinement. The flatness is then exactly what record redundancy
predicts. With the anchor present, either ratio carries the
electromagnetic-fraction correction, the quenched/unquenched ratio $S/\Sigma$ (a
\emph{quenching dual-readout}) or the photon-rich $C/\Sigma$, so starving $S$ alone
cannot break the level (the flat curve sits at the anchored value, between the
$\{\Sigma,S\}$ and $\{\Sigma,C\}$ pair values below, and the low-yield feature vector
contains no unthinned $S$ or $C$ descendant that could leak information). Measured
directly, the pair $\{\Sigma, S\}$ achieves $2.18\%$ at 20 GeV, the full dual-readout
performance, with the Cherenkov functional adding nothing further ($\{\Sigma,C\}$:
$2.37\%$; all three: $2.16\%$). At the record level any two of (unquenched sum,
quenched sum, Cherenkov path length) supply a scale and one composition ratio, the
pairings differing only in proxy quality. The photon cost appears once the unquenched
anchor is removed. The light-only record $\{S, C\}$, both thinned, the energy
estimate itself photon-limited, reaches $2.78\%$ at 20 GeV at saturating yield, a
$24$--$30\%$ premium over the anchored value across 20--100 GeV, and degrades as the
yield falls (Fig.~\ref{fig:exchange}, dashed). Where the direct Poisson term on the
sum dominates (3 pe/GeV) the measurement lands on the Gaussian pe-noise expectation
(dotted) to within a percent at 20 GeV, while at intermediate yields (30--100 pe/GeV)
it sits up to $\sim$45\% above it, photon noise amplified through the
em-fraction correction's lever arm, the noise mechanism of Ref.~\cite{Eno2025DR},
visible here as the gap between the dashed and dotted curves. The light-only
degradation is the Poisson noise of a summed functional, set by the pe-noise
quadrature and the em-fraction lever arm rather than by Prop.~\ref{prop:thin} (a
Poisson draw on a random functional is a Cox-type observation whose information does
not scale with the yield). The anchored record never feels it at the yields of
practical dual-readout crystals~\cite{CalVision2024}, and a light-only calorimeter
pays it twice, once in the sum and once in the correction.

The residual at the top level decomposes by ledger channel. Each channel's ledger truth
is regressed on the full record (truth-assisted), and its residual spread is the share
of that channel the record cannot see. Channels are species-complete, escape plus
albedo per species (backward-leaking neutrons, for instance, carry more energy than
forward-leaking ones in this geometry). Table~\ref{tab:attr} quotes all four energies.
At 20 GeV, unresolved binding is $0.69\%$, reduced by a factor $3.7$ from its $2.53\%$
marginal, the quantified success of the compensation program. In-volume neutrinos are
$0.99\%$, barely reduced from their $1.12\%$ marginal. Neutron escape-plus-albedo is
$0.51\%$, and all other escapes $0.53\%$. Quadrature closing of this budget is a
consistency diagnostic, not a decomposition identity. The channel fits are a different
estimator family (and centered widths) than the achieved-rms cell, and covariances are
not shown in the table. As a diagnostic it checks well. The diagonal quadrature sum of
the five residuals reproduces the run's achieved $\mathcal{R}_4$ rms to $-5\%$ at
20 GeV and to $-2$, $-5$ and $-10\%$ at 50, 100 and 10 GeV ($+2\%$ at 20 GeV once the
measured off-diagonal mass is added; the covariance matrix, from the same residual
vectors, carries small positive aggregate mass, $+0.28\,\%^2$ at 20 GeV, with
individual pairs $\ge -0.031\,\%^2$). The neutrino term is the sharpest design
finding. It rides on stopped-$\mu^-$ capture and decay chains, and the present record
barely tags it.

\begingroup
\squeezetable
\begin{table}[t]
\caption{Ledger attribution of the full-record residual: per-channel
truth-assisted residual spread (marginal spread in parentheses), $\sigma/E$ [\%],
from run $b$ (5--130 GeV; the canonical run for attribution and scans); channels
are species-complete (escape plus albedo); NIEL is the non-ionizing energy loss
(nuclear recoils). The quadrature sum and
that run's own achieved $\mathcal{R}_4$ give the run-consistent closing diagnostic
(see text; not an identity across estimator families);
best-of-two values are in Table~\ref{tab:ladder}. The residual covariance matrix,
computed from the same residual vectors, carries positive aggregate off-diagonal
mass at every energy ($+0.28\,\%^2$ at 20 GeV) with individual pairs at most
marginally negative ($\ge -0.031\,\%^2$).}
\label{tab:attr}
\begin{ruledtabular}
\begin{tabular}{lcccc}
channel & 10 & 20 & 50 & 100~GeV \\
\hline
neutrinos        & 1.36 (1.50) & 0.99 (1.12) & 0.60 (0.74) & 0.43 (0.57) \\
binding          & 0.98 (3.13) & 0.69 (2.53) & 0.43 (2.01) & 0.31 (1.73) \\
neutron escape   & 0.84 (1.09) & 0.51 (0.70) & 0.28 (0.40) & 0.17 (0.27) \\
other escapes    & 0.73 (0.88) & 0.53 (0.77) & 0.23 (0.29) & 0.28 (0.33) \\
NIEL             & 0.04 (0.21) & 0.03 (0.18) & 0.02 (0.15) & 0.01 (0.13) \\
\hline
quadrature sum   & 2.01 & 1.42 & 0.82 & 0.62 \\
achieved $\mathcal{R}_4$ & 2.23 & 1.48 & 0.84 & 0.66 \\
\end{tabular}
\end{ruledtabular}
\end{table}
\endgroup

The mixed-measure bound of Sec.~\ref{sec:nu} says that a record dimension which
tagged muon stops would recover the no-loss event class exactly where the
bound's hypotheses hold. Quantifying the realizable gain requires simulating such a
tag, which we have not done, but the target now has a number. The in-volume neutrino
floor is estimated at $0.99\%$ at 20 GeV, and complete removal in zero-covariance
quadrature would take the same run's $1.48\%$ to $1.10\%$, an illustrative endpoint
rather than a simulated recovery. This is a concrete and previously unquantified
argument for muon-sensitive layers or lifetime-matched delayed windows in future
hadronic calorimeters.

\subsection{What to build next}\label{sec:next}
For a dual-readout design, the measurements above order the candidate next readout
dimensions as follows. (i) Long integration (or delayed-window readout): cheap, and
measured at roughly a third of the resolution under FTFP\_BERT. (ii) Muon-stop
tagging, attacking the neutrino term that sets the estimated floor ($0.99\%$ at
20 GeV, with the illustrative endpoint $1.48 \to 1.10\%$ and its realizable gain
unmeasured). (iii) Neutron and late-composition sensitivity beyond plain late light, a
motivated candidate without a supporting number here (capture-$\gamma$ counting added
little once time structure was present). (iv) Longitudinal readout beyond
$\sim 1\lambda_I$ sampling: for single-hadron energy, last (sampling up to that
scale does carry the full shower-shape improvement, which is worth having). Items (i)
and (iv) are measured; (ii) and (iii) are candidates whose relative rank this study
does not fix. A full functional-menu ranking in the Gaussian-linear sense of
Sec.~\ref{sec:theorems} is future work, and the record-level differences above are its
coarse-grained form, usable now. The ranking is conditional on truth-level marks. It
assumes each mark is read with the record's fidelity, and realistic tag efficiency and
purity (a capture-$\gamma$ count with late-inelastic fakes; a muon tag with
punch-through mis-tags) can reorder neighboring entries. The instrument-level scan
that would settle this is the same second study as $\eta$.

\section{Implications for detector design}\label{sec:design}

\subsection{Dual-readout, recovered and extended}\label{sec:dr}
In the Gaussian-linear two-functional model, with responses normalized to the
electromagnetic scale, $S = E[f + (1-f)h_S] + n_S$, $C = E[f + (1-f)h_C] + n_C$,
$h_X \equiv (h/e)_X$ and independent channel noises, the direction in $(S,C)$ space
that annihilates the electromagnetic-fraction nuisance is unique up to scale and is
exactly the dual-readout rotation $D = (S - \chi C)/(1-\chi)$,
$\chi = (1-h_S)/(1-h_C)$, of Refs.~\cite{LeeLivanWigmans2018,WigmansRMP2018}. In our
language, the dual-readout estimator saturates the Cram\'er--Rao bound of the
two-functional record when $f$ is treated as unconstrained, with
$\mathrm{Var}(D) = 1/\Ifish_E^{\rm prof} =
(\sigma_S^2 + \chi^2\sigma_C^2)/(1-\chi)^2$ (verified to machine precision against the
profile-Fisher formula, and to $0.1\%$ against finite Monte Carlo). The observation of
Ref.~\cite{Eno2025DR} that dual-readout can hurt when channel noise dominates
the $f$-fluctuations emerges qualitatively as the crossover between two endpoint
estimators.

The framework also supplies something the fixed rotation forgoes. With an informative
prior $\sigma_f^2$ on the em-fraction spread, which is energy-dependent and
measurable, the Bayes estimator of the location model is the ridge
interpolation between the $S$-only mean-$f$ correction and the full rotation, with
location-model (frozen-covariance) Fisher information
$\Ifish_E = u^\top(C_n + \sigma_f^2 v v^\top)^{-1} u$ in closed form, where
$u = (\bar f + (1-\bar f)h_S,\ \bar f + (1-\bar f)h_C)^\top$ is the response to $E$,
$v = (1-h_S,\ 1-h_C)^\top$ the response to $f$, and $C_n = \mathrm{diag}(\sigma_S^2,
\sigma_C^2)$ the channel-noise covariance, all at $E = 1$. The marginal covariance
$M(E) = C_n + E^2\sigma_f^2 v v^\top$ contributes a further
$\tfrac12\,\mathrm{tr}[(M^{-1}M')^2]$ term, $2.7\times10^{-4}$ of the total at
$\sigma_f = 0.05$ and below $7\times10^{-4}$ over the scanned range. The ridge
estimator's variance is at or below both endpoints at every $\sigma_f$, equal
to the $S$-only endpoint where the optimal Cherenkov weight crosses zero. For
$h_S = 0.77$, $h_C = 0.21$, $\bar f = 0.55$ and
$(\sigma_S, \sigma_C) = (1\%, 3\%)$ of $E$, the gain over the better endpoint reaches
$18\%$ in variance ($10\%$ in resolution) at $\sigma_f = 0.05$, and peaks near
$28\%$ ($15\%$) at $\sigma_f \approx 0.06$. This is free resolution, using only
quantities dual-readout calorimeters already measure, with the optimal Cherenkov
weight running with energy. One never has to choose between ``ignore the
Cherenkov channel'' and ``apply the full rotation''. The criterion of
Ref.~\cite{Eno2025DR} marks the crossing of two suboptimal endpoints.

Adding any third functional, a neutron time-tail fraction, a capture-$\gamma$ count,
or a wavelength bin, appends a row to the response matrix, with closed-form marginal
Fisher gain in the Gaussian model and the empirical Bayes-risk ranking of
Sec.~\ref{sec:theorems} as its general surrogate. Its coarse-grained measured form is
the record-level ordering and candidate ranking of Secs.~\ref{sec:ladder} and
\ref{sec:next}.

\subsection{Information efficiency of record archetypes}
Taking the best achieved value across record levels at each energy as the reference,
the achieved-risk ratio
$\hat\eta = (\sigma_{\rm ref}/\sigma_{\mathcal{R}})^2$ of each record family is the
measurable estimate of the information efficiency $\eta$, a ratio of two achieved
values rather than of two floors. It follows directly from Table~\ref{tab:ladder}
(computed from unrounded values), quoted to the precision the cross-run stability
supports, with the reference row definitionally at 1:

\begin{ruledtabular}
\begin{tabular}{lcc}
record family & $\hat\eta$ (20 GeV) & $\hat\eta$ (100 GeV) \\
\hline
sum only                       & 0.10 & 0.06 \\
+ dual-readout ($\mathcal{R}_1$) & 0.36 & 0.46 \\
+ shower shape                 & 0.65 & 0.78 \\
+ time / full record           & $\ge 0.96$ & 1 (ref.) \\
\end{tabular}
\end{ruledtabular}

\medskip
\noindent A classical sum calorimeter, however perfectly built, extracts
$\lesssim$10\% of what the full-record estimator achieves on the same showers. The
anchored dual-readout level recovers a third to a half. The time- and count-enriched
records are statistically indistinguishable from the full-record reference within the
measured estimator scatter. These are statements about achieved risks within this
record family, since the reference is itself an achieved value, and they are computed
on truth-level records. Instrument $\eta$ (optical transport, digitization, noise)
multiplies on top and is the natural second study. The exchange-rate curves of
Fig.~\ref{fig:exchange} state what a nanosecond of gate, a $\lambda_I$ of
segmentation, or a decade of light yield is worth, at one fixed, validated estimation
procedure rather than per reconstruction algorithm.

\subsection{Implications for future programs}
The comparisons detector proposals need are the lattice-monotone quantities this
framework defines, floors per record specification, exchange rates per readout
dimension, and $\eta$ per design, decoupled from the reconstruction algorithm of the
day. That decoupling is the property optimization
frameworks~\cite{Chung2026Agentic} require of an objective. What is measured here are
their achieved-risk estimates ($\hat\eta$, record-level values, scan curves) at a
fixed, validated procedure: the operational stand-ins, labeled as such, until sharper
bounds exist. Hardware moves the record; algorithms approach its floor. The two claims
have separate, measurable meanings.

\section{Limitations and outlook}\label{sec:outlook}
The values reported here are conditional on the cascade model (Geant4 FTFP\_BERT), on
the record specifications stated, and, for achieved values, on the feature families
stated, with the cost of withheld sufficient statistics measured in
App.~\ref{app:protocol}. The physics-list sensitivity is measured directly, in two
directions. Applying the FTFP\_BERT-trained estimators to QGSP\_BERT events at 20 GeV
shifts the achieved resolution by $+0.7\%$ relative on the dual-readout record and
$+19\%$ on the full record.

We also comment on the neutron transport model. This study's production uses FTFP\_BERT's
standard (non-HP) neutron physics with the tracking cut removed. The stock
high-precision reference list FTFP\_BERT\_HP differs from it in three ways, not one:
the ENDF-based neutron transport below 20 MeV, an additional radioactive-decay
constructor (\texttt{G4RadioactiveDecayPhysics}, absent from FTFP\_BERT) that decays
every unstable residual nucleus whose sampled decay time is below one year and adds
its $\beta$ and $\gamma$ energy to the infinite gate, and a proton production cut of
zero in place of 0.7~mm. A $2\times2$ set of 12k-event 20 GeV productions with
identical seeds separates the three (App.~\ref{app:g4}). With the stock list, 222
radioactive decays per event release 324 MeV per event. By the controlled decay
toggle, $190$ MeV of that goes to neutrinos and $118$ MeV is deposited after
10~$\mu$s, with the balance deposited promptly, and this decay energy is the dominant
source of the apparent $\times 5$ increase in light after 10~$\mu$s and $\times 1.6$
increase in in-volume neutrino energy that the stock list shows relative to the
baseline. With the decay constructor removed and the cuts matched, the neutrino energy
is unchanged ($\times 1.00$) and the light after 10~$\mu$s rises by $\times 1.36$. The
neutron-transport differences proper are therefore: $13.5\%$ fewer captures, a $59\%$
larger fraction of capture $\gamma$s arriving after 10~$\mu$s ($37\%$ more in count),
$\times 1.36$ the light after 10~$\mu$s, $\fem$ and the binding fraction unchanged
within $0.5\%$, and a total-deposit spread larger by $6\%$ (the proton cut alone moves
it by $2.6\%$).

Transferring the FTFP\_BERT-trained estimators (run $a$, the procedure of
Sec.~\ref{sec:ladder}) to the clean high-precision test sample shifts them by $+6\%$
(dual-readout record), $+47\%$ (+time), and $+47\%$ (full record), relative. Those
shifts mix cascade-model differences with feature-distribution shift, and only
retrained estimators separate them. Table~\ref{tab:hp} lists the 20 GeV values
retrained and evaluated entirely within the clean high-precision configuration,
against the baseline and the stock list. The dissipated-energy sum has the largest
absolute shift under the transport-model change ($+7.4\%$ relative, the counterpart of
the $6\%$ larger deposit spread), the dual-readout and shower-shape levels move by
$1\%$ or less, the time level by $+8.9\%$, and the full record by $+0.4\%$, inside the
baseline cross-run spread of $0.19\%$ and the high-precision pair's own $0.18\%$ (in
absolute $\sigma/E$). The resolution gained by adding time structure to the
shower-shape record is $0.17\%$ under high-precision transport ($0.18$ and $0.17\%$ in
the two runs) against $0.28\%$ under FTFP\_BERT, and the counting marks recover part
of the difference ($0.14\%$ against $0.03\%$ from $\mathcal{R}_3$ to $\mathcal{R}_4$).
Standard neutron transport therefore overstates the value of late-time structure at
20 GeV by a factor of about $1.6$. The stock list had suggested a factor of three
(Table~\ref{tab:hp}, last column; time increment $0.09\%$). The difference is the
decay energy, a delayed proxy of spallation activity that flattered the stock-list
dissipated-energy sum and compressed its time increment. The same-events
one-functional anchor moves the other way, to $2.08\%$ on the clean sample, so the
full record sits a factor $1.6$ below it under high-precision transport as well, and
the transferred shifts were therefore mostly feature-distribution shift.

\begin{table}[t]
\caption{Achieved $\sigma/E$ [\%] (rms) at 20 GeV per record level under three
physics configurations: the baseline FTFP\_BERT ladder (best of two runs,
Table~\ref{tab:ladder}); the ladder retrained and evaluated entirely within the clean
high-precision configuration (FTFP\_BERT\_HP with the radioactive-decay constructor
removed and production cuts matched; two independent 200k-event runs mirroring the
baseline pair, disjoint seeds and spans, 12k-event test set); and the stock
FTFP\_BERT\_HP list (radioactive decay and zero proton cut included), whose late
sector is dominated by the decay energy (text). Cross-run rms spreads at 20 GeV on
the full record: $0.19\%$ (baseline) and $0.18\%$ (clean HP), in absolute
$\sigma/E$.}
\label{tab:hp}
\begin{ruledtabular}
\begin{tabular}{lccc}
record & FTFP\_BERT & clean HP & stock HP \\
\hline
$\mathcal{R}_0$ sum        & 4.07 & 4.37 & 3.99 \\
$\mathcal{R}_1$ +$S,C$     & 2.16 & 2.15 & 2.21 \\
$\mathcal{R}_2$ +shape     & 1.60 & 1.61 & 1.52 \\
$\mathcal{R}_3$ +time      & 1.32 & 1.44 & 1.43 \\
$\mathcal{R}_4$ +counts    & 1.29 & 1.30 & 1.42 \\
\end{tabular}
\end{ruledtabular}
\end{table}

The structure of the sensitivity is the statement that the increments and
the intermediate record levels are model-conditional; the top, at this energy and
between the two retrained transport models, is not. The time level inherits the
uncertainty of low-energy neutron transport, because late-time and multi-channel
correlations are exactly where generators disagree, the dissipated-energy sum inherits
it through the deposit spread, and the full record recovers from one dimension what it
loses in another. Test-beam anchoring of those correlations (DREAM time
structure~\cite{Akchurin2009Neutron}, CALICE imaging and
timing~\cite{CALICE2013G4Val,CALICET3B2014}) is correspondingly most valuable for the
intermediate levels and the increments between them. These are single-particle
results. Jets, pileup, and particle-flow contexts change the estimand and the record,
not the framework, and are the natural second paper, as is the instrument-level $\eta$
study of realistic detector concepts with full optical transport and digitization. The
deepest model dependence, the nuclear physics of binding-energy release and its
correlates, is shared with every compensation and dual-readout study since the 1980s.
The framework at least makes it explicit, and test-beam data (DREAM/RD52 time
structure~\cite{Akchurin2009Neutron}, CALICE shower imaging~\cite{CALICE2013G4Val} and
time structure~\cite{CALICET3B2014}) can calibrate the relevant mark distributions
where simulation is weakest.

\medskip
\noindent\emph{Code and data availability.} The Geant4 energy-ledger application
(\texttt{g4ledger}), every production macro and seed, the analysis and figure code,
and the merged numerical results for each table and figure (including the per-run
intermediates behind the best-of selections) are archived together with the per-event
production data (with SHA-256 checksums) in the project repository, available from the
author upon request.

\begin{acknowledgments}
The ideas and concepts for the framework presented in this work were solely conceived
by the author, with simulation setup, execution, analysis, and review assisted by the AI systems
Claude (Fable~5 and Fable~5.1, Anthropic) and GPT-6 Astra (OpenAI). The author directed 
and refined the study at all stages. The author reviewed and verified all code and outputs
and takes sole responsibility for the scientific content of the paper.
\end{acknowledgments}

\appendix

\section{Proofs}\label{app:proofs}

\subsection{Proposition~\ref{prop:dpi}}
Fisher-information data processing for Markov kernels independent of the parameter is
Ref.~\cite{Zamir1998}, and mutual-information monotonicity is the classical DPI.
Record refinements in the lattice of Sec.~\ref{sec:framework} (coarser cells, shorter
gates, dropped marks, channel subsets) are deterministic functions of the finer
record, hence kernels.

\subsection{Proposition~\ref{prop:thin}}
For a Poisson point process with intensity $\lambda_E(s)$ on $\mathcal S$, the
log-likelihood of a realization $\{s_i\}$ is
$\sum_i \log\lambda_E(s_i) - \int_{\mathcal S}\lambda_E$, so the score is
$\sum_i \partial_E \log\lambda_E(s_i) - \int \partial_E\lambda_E$ and, by Campbell's
formula and Poisson variance,
$\Ifish(E) = \int_{\mathcal S} (\partial_E\lambda_E)^2/\lambda_E$. Independent
retention with probability $p(s)$, independent of $E$, yields a Poisson process with
intensity $p\lambda_E$, whence
$\Ifish_{\rm thin}(E) = \int p\,(\partial_E\lambda_E)^2/\lambda_E$; constant $p$
factors out. $\square$

\subsection{Proposition~\ref{prop:cluster}}
With $N \sim \mathrm{Poisson}(\mu)$, masses i.i.d.\ with moments $m_1, m_2$ and
$\mathrm{cv}^2 = m_2/m_1^2 - 1$, the sum $S$ has cumulants $\kappa_r = \mu\,m_r$
($m_r$ the $r$-th mass moment). Replacing the density of $S$ by its Gaussian surrogate
(matched mean and variance),
\[
\Ifish_S(\mu) \simeq \frac{(\mathrm{d}\bar S/\mathrm{d}\mu)^2}{\mathrm{Var}\,S}
 + \frac{(\mathrm{d}\,\mathrm{Var}S/\mathrm{d}\mu)^2}{2\,\mathrm{Var}^2 S}
 = \frac{1}{\mu}\,\frac{1}{1+\mathrm{cv}^2} + \frac{1}{2\mu^2},
\]
whose leading term is the proposition's statement and whose second term is the
surrogate's variance-derivative contribution, not the true finite-$\mu$ correction
(the third cumulant enters at the same order; for gamma masses the exact correction is
several times smaller). A local limit theorem for the density of $S$ and for its
$\mu$-derivative, available under the stated smoothness hypothesis, is what licenses
the leading term. Uniform integrability of the score holds unconditionally, since the
observed score is $\mathbb{E}[N/\mu - 1 \mid S]$, and the exact identity
$\Ifish_S = 1/\mu - \mathbb{E}[\mathrm{Var}(N \mid S)]/\mu^2$ bounds every mass law by
the cluster-resolved value. The density local limit does not apply to discrete mass
laws. For count-identifying laws, such as two masses in an irrational ratio, $S$
determines $N$ and $\Ifish_S = 1/\mu$ ($N$ is sufficient; the mass distribution
carries no $\mu$-dependence), while lattice support alone does not imply count
identification. Corrections beyond normal order involve the higher standardized
cumulants of $S$, which scale as $\mu^{-r/2+1}$. The exact $\Ifish_S(\mu)$ used as
numerical reference is obtained from the characteristic function
$\varphi_S(u) = \exp[\mu(\varphi_M(u)-1)]$ by FFT inversion of the density family and
quadrature of $(\partial_\mu f)^2/f$. Verified values are quoted in
Sec.~\ref{sec:cluster}. $\square$

\subsection{Proposition~\ref{prop:stam}}
Left inequality: Stam's superadditivity of inverse Fisher information under
convolution of independent densities~\cite{Stam1959}, equality iff all factors
Gaussian. Right inequality: for any density with finite variance, $J\,\sigma^2 \ge 1$
(Cram\'er--Rao/Cauchy--Schwarz on the location score), applied to the convolution,
whose variance is $\sum_k\sigma_k^2$; equality again characterizes the Gaussian.
$\square$

\subsection{Lemma~\ref{lem:nu} and the mixed-measure bound}
Under the stated factorization the family satisfies
$p(\rec \mid E) = \int p_0(\rec \mid E-u)\, f_\nu(u)\,\mathrm{d}u$: the parameter
enters only through the location-smoothed first stage. By Prop.~\ref{prop:dpi} applied
to the chain $E \to E-\nu \to \rec$,
$\Ifish(E;\rec) \le \Ifish(E;\,E-\nu) = J(f_\nu)$. For the mixed case, pass first to
the smoothed variable $V = E - \nu$, through which alone the record depends on $E$
under the factorization, so that $\MMSE(E \mid \rec) \ge \MMSE(E \mid V)$ by
Prop.~\ref{prop:dpi}; then condition on the loss count $D$, which can only decrease
the Bayes risk, so with prior $\pi$,
\begin{align*}
&\MMSE(E \mid \rec)\;\ge\;\MMSE(E \mid V)\\
&\quad\ge\;\sum_d P(D{=}d)\;\MMSE\big(E \mid V, D{=}d\big)\\
&\quad\ge\; \sum_{d\ge 1} \frac{P(D{=}d)}{\,J(f_{\nu|d}) + \Ifish(\pi)\,},
\end{align*}
using the van Trees inequality within each conditional experiment, in which $V$ is a
location family with kernel $f_{\nu|d}$ ($\Ifish(E; V \mid d) = J(f_{\nu|d})$;
$\Ifish(\pi)$ the prior's own Fisher information; the $d{=}0$ term vanishes as
$J \to \infty$, consistent with exact recoverability of the no-loss class). The
passage through $V$ makes the per-class step valid: conditioning the record
itself on $D$ would need the stronger hypothesis $\rec \perp E \mid (V, D)$, because a
mark independent of the loss unconditionally can identify it within a class. The
display carries two idealizations: The loss count is treated as
energy-independent across the prior support (otherwise $\pi$ is replaced by the tilted
conditional prior $\pi_d \propto \pi \, P(D{=}d \mid E)$, whose own information enters
the denominator), and each $f_{\nu|d}$ is taken as an $E$-independent location kernel,
as in the Lemma. Both drift in real cascades, where loss counts grow with energy, so
the display is the skeleton of the mixed-case bound under stated conditions, not its
sharpest physical form. The Geant4 sections use the estimated neutrino floor, never
this display, for quantitative statements. The inequality is stated for priors with
vanishing boundary terms (smooth-edged; a hard-edged log-uniform prior is used only
for training in the measurements, where all quoted evaluations are
prior-interior). For broad priors $\Ifish(\pi)$ is negligible against $J(f_{\nu|d})$.
$\square$

\section{Geant4 bookkeeping and energy balance}\label{app:g4}

The accounting assigns every track a single ledger energy $L$: kinetic energy
for baryons and nuclei (baryon number is conserved; rest mass never becomes shower
energy), total energy for mesons and $\mu/\tau$ (their mass is live energy, released
at decay or capture), $E_{\rm kin}$ for $e^-$, $E_{\rm kin} + 2m_e$ for $e^+$
(annihilation potential), analogously for antibaryons. With this convention every
electromagnetic vertex and the dominant meson decay modes ($\gamma\gamma$, $\mu\nu$,
$\pi\pi$) balance identically (electron semileptonic modes carry $\pm m_e$), and for
every step the kinetic-ledger imbalance
\[
  q_{\rm step} = L_{\rm pre} - L_{\rm post} - e_{\rm dep}
  - \textstyle\sum_{\rm sec} L
\]
telescopes over the event to the exact identity
$E_{\rm ref} = \sum e_{\rm dep} + \sum L_{\rm esc} + \sum q$. The energy balance is
therefore a conservation identity, not a fit. Measured residuals are at
double-precision rounding ($\mathrm{mean}|c|/E = 8.7\times10^{-15}$ for $e^-$ at
10 GeV and $8.5\times10^{-14}$ for $\pi^-$ at 20 GeV in the 100-event acceptance runs;
thresholds of $10^{-3}$/$5\times10^{-3}$ cleared by eleven orders; the 12k-event
production samples give $\mathrm{mean}|c|/E = 2.6\times10^{-13}$, raised by the
upstream-decay events noted below). Positive $q$ accumulates gross
binding/rearrangement loss (the $\Delta B$ of Ref.~\cite{LeeLivanWigmans2018}, with
double counting excluded by construction); negative $q$ collects binding
released (capture $\gamma$s at $-S_n \simeq -7.6$~MeV per capture in Cu,
strangeness-mass release); sub-threshold dust is carried explicitly in a residual
column and enters the balance.

Datasets: G4NDL 4.7.1, PhotonEvaporation 6.1.2, EMLOW 8.8, PARTICLEXS 4.2. The
capture-$\gamma$ and low-energy-neutron content of the record depends on these
versions. Physics-list control (Sec.~\ref{sec:outlook}): the application can remove
the radioactive-decay constructor from a reference list and keep the macro's
production cuts for all four particles (two opt-in switches), and the $2\times2$ set
of 12k-event 20 GeV productions, radioactive decay on/off and proton cut 0/0.7~mm with
the seeds of the HP test set, is what separates neutron transport from the two other
differences of the stock FTFP\_BERT\_HP list. The per-process $Q$ table shows the
decays directly (222 per event releasing 324 MeV at 20 GeV with the stock list, none
with FTFP\_BERT). Geant4-specific findings are handled at the source. FTFP\_BERT's
default neutron tracking cut kills neutrons at 10~$\mu$s and silently discards their
kinetic energy and late captures ($\sim$2\% of capture $\gamma$s arrive later); it is
removed. The bundled \texttt{G4GammaGeneralProcess} wrapper destroys per-process
attribution; it is disabled, after the physics-list constructor, which otherwise
re-enables it. Generator-level vertex non-conservation lands inside the measured $q$
of its vertex, absorbed into the nuclear term where it belongs. Escaping tracks are
recorded and then killed at the boundary, making escape single-entry. The monitor for
energy deposited in the world volume stayed at zero throughout, and the monitor for
ledger energy born in the world volume is non-zero for 15 of the 448k primaries
($3\times10^{-5}$): 14 that decayed to $\mu\nu$ in the 5~cm vacuum gap between the gun
and the block, punch-through muons rather than showers (in two of which the muon stops
in the block; their ledgers balance), plus the boundary-birth corner counted
separately below. Every evaluation sample in this paper applies the pre-specified cut
$l_{\rm worldborn} > 0$, the simulation's equivalent of an upstream veto. It removes
one event from each of the 10 and 20 GeV baseline test sets and one from each 20 GeV
physics-list test sample, and the training samples are left as produced. The 20 GeV
event, a muon that punched through, had shifted the $\mathcal{R}_1$ value of
Table~\ref{tab:ladder} by $0.08\%$ and its $\mathcal{R}_4$ value by $0.02\%$ with the
same estimators (absolute $\sigma/E$); at 10 GeV the shifts were $0.01\%$ and below.
Per-event master-drawn seeds make merged output identical across thread counts. Across
the full 448k-event $\pi^-$ production the largest single-event
$|\mathrm{residual}|/E$ is $1.0\times10^{-5}$, one event, a boundary-birth corner case,
excluded (and counted) by the analysis loader's $10^{-6}$ requirement; the runner-up
is $5\times10^{-9}$. One escape convention matters for interpreting the attribution.
An exiting muon is ledgered at its total energy in the muon-escape column, so
neutrinos from decays outside the volume are counted as muon escape, not
neutrino loss; the $\nu$ channel is in-volume production only. Cross-checks that
balance arithmetically: late ($>$10 ns) energy in $e^-$ showers equals the
capture-count $\times S_n$ product (approximately: the effective $S_n$ is isotope- and
cross-section-weighted); $e^-$ showers in Cu spend $0.23\%$ of $E$ on photonuclear
binding at 10 GeV and $0.30\%$ at 50 GeV; $\pi^-$ at 20 GeV sends $1.5\%$ of $E$ into
neutrinos (species-complete, in-volume; dominated by stopped-$\mu$ chains) and $8\%$
of its visible energy arrives later than 10~ns. In a documented sanity run (output not
archived) Pb roughly doubles the neutron yield relative to Cu (828 against 445 created
per event) and raises the binding fraction by a quarter (0.179 against 0.141), as
expected from $Z$-scaling. Full conventions, the per-process $Q$ table, and the
acceptance log are documented with the code.

\section{Estimator validation}\label{app:protocol}
The license for quoting trained-estimator residuals as upper brackets of floors is
calibration on cascades where the floor is exactly computable. There are four stages,
with all acceptance criteria listed in the analysis repository.

\emph{(i) Exactly solvable cascade model.} A generative hadronic cascade model (Beta
em-fraction $\to$ Poisson vertices $\to$ Gamma binding $\to$ Poisson neutron tag
$\to$ compound-Poisson untagged losses with an atom at zero) admits an exact posterior
by $(E,B)$ quadrature. Self-tests of the exact posterior: on prior-drawn ensembles,
where the Bayes identity
$\langle (E-\hat E_{\rm post})^2 \rangle = \langle \mathrm{Var}_{\rm post}\rangle$
holds, the consistency ratio
$\mathrm{rms}(E-\hat E_{\rm post})/\sqrt{\langle \mathrm{Var}_{\rm post}\rangle}$
reads $0.999$--$1.008$ across record levels, binned conditional calibration is
$0.98$--$1.03$ across record quintiles, and the probability-integral-transform (PIT)
Kolmogorov distance is $0.007$--$0.013$ at $N = 16$k (consistent with uniformity at
$\mathcal{R}_2$--$\mathcal{R}_3$, $p = 0.4$ and $0.1$; the $\mathcal{R}_1$ value
reflects the fast moment-matched vertex marginal and reads $0.007$ with the exact
marginal). On fixed-$E$ ensembles, where the identity need not hold, the same ratio
reads $0.984$--$1.012$. The exact vertex-marginal posterior and the fast
moment-matched form agree to $2\times10^{-4}$ relative in width. The estimator test:
gradient-boosted trees and multilayer perceptrons trained across a log-uniform
spectrum, evaluated on fixed-$E$ ensembles on identical events with identical
metrics, reach a best-to-exact-posterior rms ratio of $1.0000$--$1.0034$ (paired
bootstrap uncertainties $\le 0.8\times10^{-3}$) at every record level and energy. The
procedure recovers the exact posterior's rms when the record's sufficient statistics
are available to it.

\begin{figure}[t]
\includegraphics[width=\columnwidth]{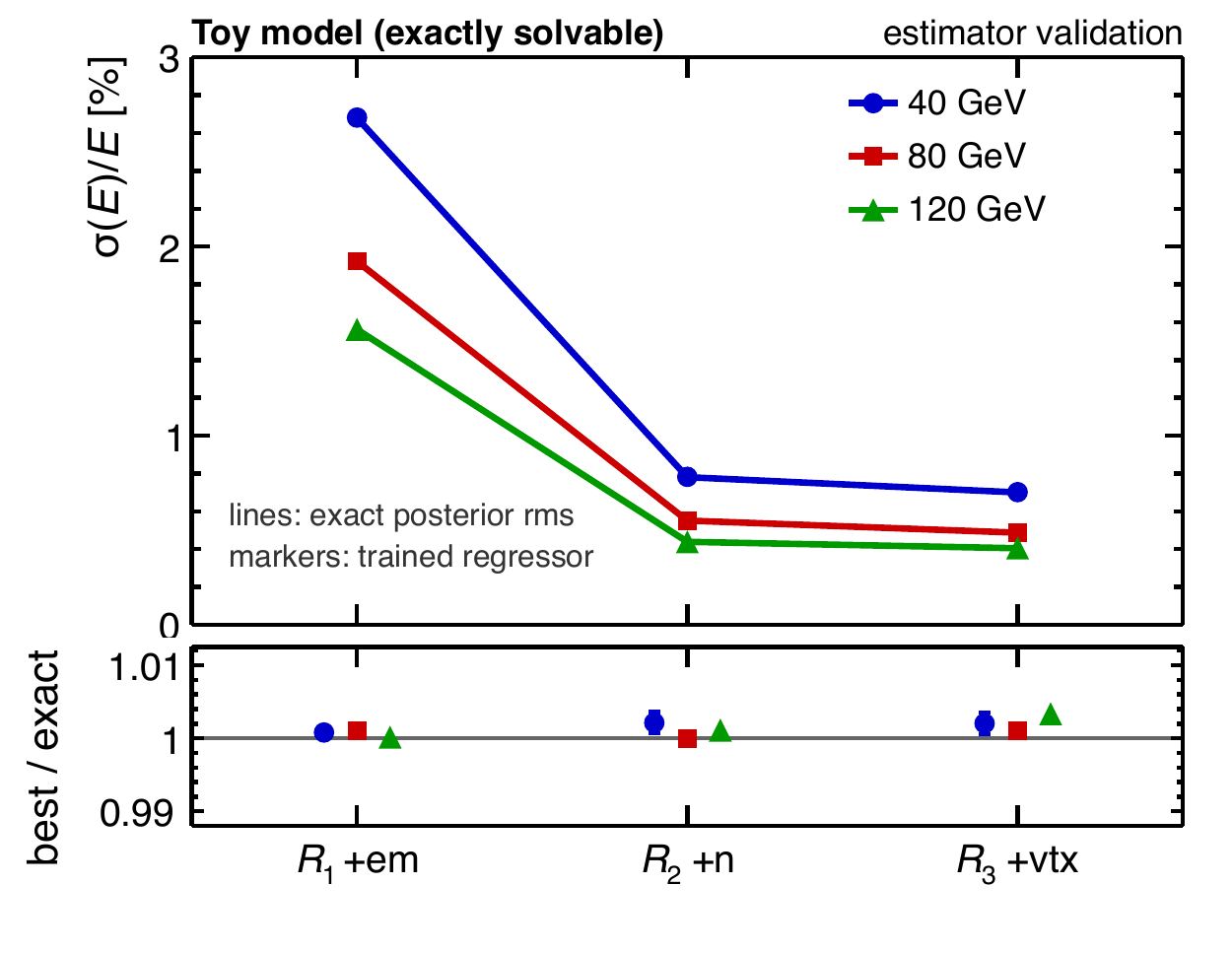}
\caption{Estimator validation on the exactly solvable cascade model. Top: the exact
posterior mean's rms error on fixed-energy ensembles (lines; $\mathcal{R}_1$--$2$
through the moment-matched vertex marginal, cross-checked against explicit
marginalization) and trained-regressor performance (markers) per record level at three
fixed test energies. Bottom: the paired best-regressor to exact-posterior ratio with
$2\sigma$ bootstrap uncertainties. The procedure recovers the exact posterior's rms to
a few parts in $10^3$ at every record level and energy.}
\label{fig:protocol}
\end{figure}

\begin{figure}[t]
\includegraphics[width=\columnwidth]{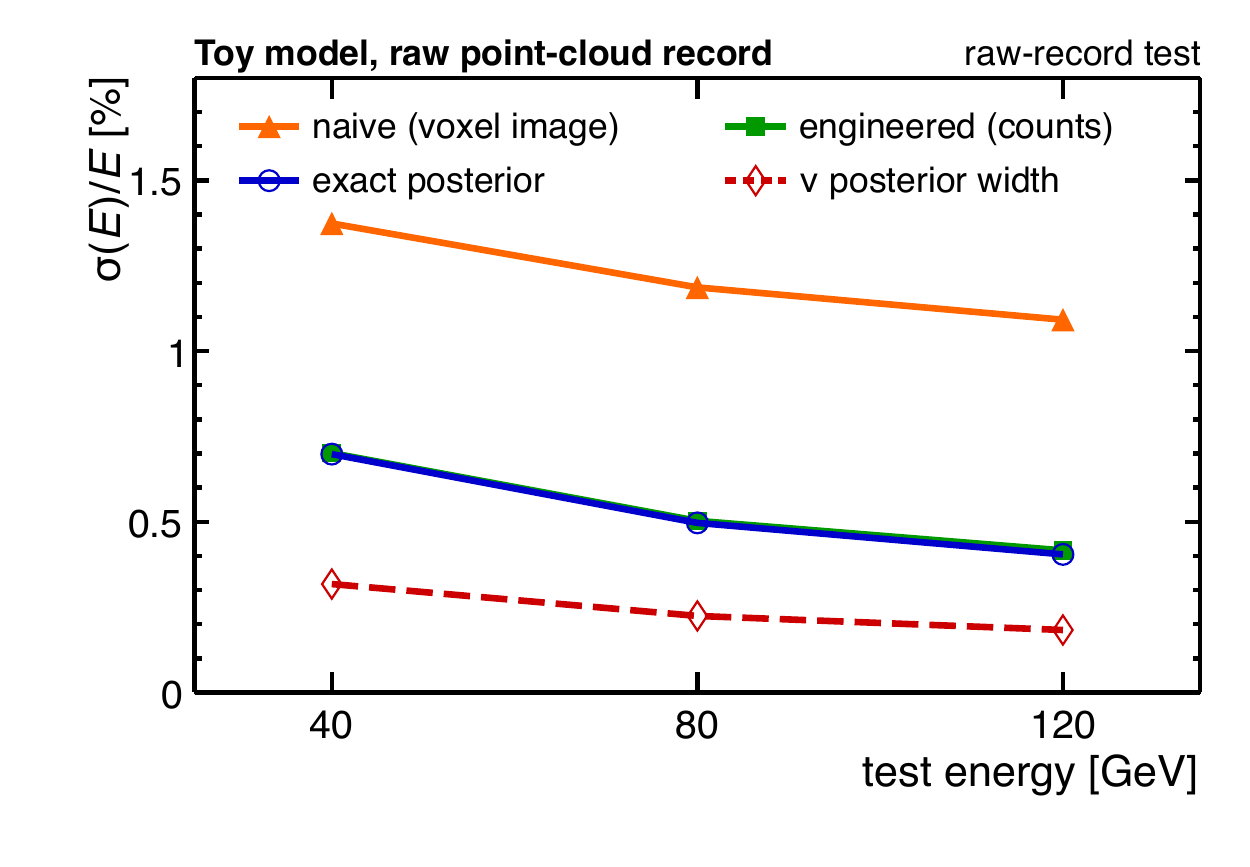}
\caption{Raw-record control test: the same cascade exposed as an unordered point
cloud, the engineered feature set additionally retaining the true zero-vertex
indicator (its exact-posterior benchmark uses that augmented record). Engineered
counting features (filled squares) recover the exact posterior's rms (open circles) to
$0.4$--$2.9\%$ relative; a naive voxel-image feature set pays a factor $2.0$--$2.6$;
dashed: the posterior width of the untaggable channel,
$\sqrt{\langle \mathrm{Var}(E \mid V, B, E_{\rm em}, N_v)\rangle}/E$ with the binding
loss supplied (a diagnostic, not a bound on the fixed-energy rms), which no feature
set undercuts.}
\label{fig:toy3b}
\end{figure}

\emph{(ii) Raw-record control test.} The same latent physics is exposed as an unordered
point cloud (per-vertex tuples, an energy-carrying prompt tuple when no vertex fires,
and zero-energy late hits with ancillary positions, times, and junk marks). Sufficient
statistics are not hand-fed, with one stated exception: the engineered feature set
retains the true zero-vertex indicator, and its exact-posterior benchmark uses that
augmented record. This measures the cost of withheld sufficient statistics. A naive
``voxelized-image'' feature set degrades by a factor $2.0$--$2.6$ (worst at high
energy, where occupancy saturation of the late-hit grid erodes counting information,
an emergent miniature of real detector occupancy) relative to engineered counting
features, which recover the exact posterior to $0.4$--$2.9\%$ relative, and no feature
set falls below the untaggable-channel posterior width. Geant4 values are accordingly
quoted as achieved with a stated feature family, with convergence evidence, not
as assumption-free infima.

\emph{(iii) Application checks.} On Geant4 output: event-level energy balance
(App.~\ref{app:g4}), reproduction of the one-functional conditional-variance value of
Ref.~\cite{LeeLivanWigmans2018} before any extension, feature/statistics convergence
scans, physics-list variation, and a tail-conditional check. On the
charge-exchange-like $\fem > 0.9$ subpopulation (1.2--1.6\% of events), the class on
which the linearized ratio construction diverges, the full-record estimator's residual
rms is a factor $\sim 2$ smaller than its ensemble value ($0.79$ vs. $1.48\%$
at 20 GeV within the wide run) with bias $\le 0.26\%$. EM-like events carry little
invisible energy, and the record-level estimator handles the class the ratio
construction cannot.

\emph{(iv) Selection rule.} Quoted values are the smallest test-set rms across two
regressor families (gradient-boosted trees with a 5\% early-stopping holdout;
multilayer perceptron with internal validation) and two independent training runs
(log-uniform 8--120 and 5--130 GeV, disjoint seeds from all test runs; seeds and spans
fixed in advance in the run configurations). Selecting on the test metric is itself
validated. On the cascades the matching two-family model selection (settings
differ in the holdout) returns best-to-exact ratios of $1.0000$--$1.0034$, and
selecting the family on one ensemble while evaluating it on an independent one returns
$0.9997$--$1.0034$. The selection-on-metric optimism is thereby measured directly,
$-0.0009$ to $+0.0015$ per cell with mean $+0.0002$. The run-level best-of is a
second, separate selection, not calibrated by the cascade model, and its scale is measured
empirically by the cross-run spreads. The 12k-event test rms itself carries
$\approx 0.6\%$ relative statistical error for Gaussian residuals and more with the
observed tails, tail-driven and cell by cell. Several of the spreads quoted sit below
that error, and a small spread there is an agreement observation between the two training runs, not a
precision statement. The training-lottery variability
across independent runs is a distinct effect, quantified directly by the cross-run
spreads of App.~\ref{app:runs} (part of the 10 GeV spread reflects training-span edge
proximity: the two runs' lower edges sit at 8 and 5 GeV). Each run's value bounds the
floor from above by itself. The best-of is simply the tighter bound, and no
record-level difference smaller than the printed spread is interpreted. Per-family
values are published alongside the merged results in the analysis data products.

\section{Both-runs ladder values}\label{app:runs}

Table~\ref{tab:runs} lists the achieved values of both independent end-to-end
training runs behind every best-of cell of Table~\ref{tab:ladder}, with the
absolute cross-run rms spread $|\Delta|$ quoted as the stability systematic.

\begin{table}[t]
\caption{Per-run achieved $\sigma/E$ [\%] (rms, robust width in parentheses):
run $a$ trained on log-uniform 8--120 GeV, run $b$ on 5--130 GeV, disjoint seeds
throughout.}
\label{tab:runs}
\begin{ruledtabular}
\begin{tabular}{llcccc}
record & run & 10 & 20 & 50 & 100~GeV \\
\hline
$\mathcal{R}_0$ & $a$ & 5.09 (4.96) & 4.07 (3.83) & 3.05 (2.86) & 2.58 (2.42) \\
 & $b$ & 5.12 (4.86) & 4.11 (3.84) & 3.06 (2.95) & 2.59 (2.52) \\
 & $|\Delta|$ & 0.03 & 0.04 & 0.01 & 0.02 \\
$\mathcal{R}_1$ & $a$ & 2.85 (2.57) & 2.16 (1.79) & 1.27 (1.13) & 0.97 (0.81) \\
 & $b$ & 3.07 (2.70) & 2.19 (1.80) & 1.27 (1.13) & 0.97 (0.81) \\
 & $|\Delta|$ & 0.22 & 0.03 & 0.00 & 0.00 \\
$\mathcal{R}_2$ & $a$ & 2.39 (2.10) & 1.60 (1.43) & 0.97 (0.89) & 0.74 (0.62) \\
 & $b$ & 2.44 (2.15) & 1.66 (1.49) & 0.98 (0.91) & 0.74 (0.60) \\
 & $|\Delta|$ & 0.05 & 0.07 & 0.01 & 0.00 \\
$\mathcal{R}_3$ & $a$ & 2.17 (1.77) & 1.32 (1.17) & 0.85 (0.75) & 0.66 (0.50) \\
 & $b$ & 2.12 (1.86) & 1.41 (1.23) & 0.87 (0.77) & 0.67 (0.51) \\
 & $|\Delta|$ & 0.05 & 0.09 & 0.02 & 0.01 \\
$\mathcal{R}_4$ & $a$ & 2.60 (1.78) & 1.29 (1.12) & 0.81 (0.66) & 0.66 (0.53) \\
 & $b$ & 2.23 (1.83) & 1.48 (1.23) & 0.84 (0.75) & 0.66 (0.51) \\
 & $|\Delta|$ & 0.37 & 0.19 & 0.03 & 0.00 \\
\end{tabular}
\end{ruledtabular}
\end{table}

\bibliography{refs}

\end{document}